\documentclass[11pt,a4paper]{article}

\usepackage{graphicx}
\usepackage{amssymb}
\usepackage{amsmath}
\usepackage{amsfonts}
\usepackage{dsfont}
\usepackage{mathtools}
\usepackage{array}
\usepackage{comment}
\usepackage{rotating}
\usepackage{bbold,amsfonts}
\allowdisplaybreaks

\usepackage[utf8]{inputenc}
\usepackage{bm}
\usepackage{xcolor}
\usepackage{float}
\usepackage{braket}
\usepackage{placeins}
\usepackage[height=8.8in,width=6.45in]{geometry}
\usepackage[font=small,labelfont=bf]{caption}
\usepackage[hidelinks]{hyperref}
\usepackage{booktabs,float,slashed}
\usepackage{standalone}
\usepackage{caption}
\usepackage{subcaption}
\usepackage{makecell}

\usepackage[maxbibnames=99,sorting=none,giveninits=true,backend=biber,style=numeric-comp,sortcites,doi=false,hyperref=true]{biblatex}
\renewbibmacro{in:}{}
\usepackage{hyperref}
\DeclareFieldFormat{doilink}{\iffieldundef{doi}{#1}{\href{https://doi.org/\thefield{doi}}{#1}}}
\DeclareFieldFormat[article,periodical]{volume}{\mkbibbold{#1}}
\DeclareFieldFormat[article,periodical]{journaltitle}{#1}
\DeclareFieldFormat[article,periodical]{pages}{#1}

\usepackage{xpatch}
\xpatchbibdriver{article}
  {\usebibmacro{journal+issuetitle}%
   \newunit
   \usebibmacro{byeditor+others}%
   \newunit
   \usebibmacro{note+pages}}
  {\printtext[doilink]{%
     \usebibmacro{journal+issuetitle}%
     \newunit
     \usebibmacro{byeditor+others}%
     \newunit
     \usebibmacro{note+pages}}}
  {}{}

\numberwithin{equation}{section}

\newcommand{\beq}{\begin{equation}}
\newcommand{\eeq}{\end{equation}}

\DeclareMathOperator{\Tr}{Tr}
\DeclareMathOperator{\tr}{tr}

\newcommand{\ii}{\mathrm{i}}

\makeatletter
\newcommand*{\letterdef@}{}
\newcommand*{\letterdef}[3]{%
	\def\letterdef@##1{\expandafter\newcommand\csname #1\endcsname{#2{##1}}}%
	\@tfor\@tempa :=#3\do{\expandafter\letterdef@\expandafter{\@tempa}}}
\makeatother
\letterdef{c#1} {\mathcal}{ABCDEFGHIJKLMNOPQRSTUVWXYZ}
\letterdef{rm#1}{\mathrm} {dDeimM}

\newcommand{\E}{{\scriptscriptstyle{\mathbf{E}}}}
\newcommand{\D}{{\scriptscriptstyle{\mathbf{D}}}}
\begin{document}

\begin{titlepage}

\vspace*{10mm}
\begin{center}
{\LARGE \bf 
Bremsstrahlung function in $\mathcal{N}=2$ SCFTs 
\\[2mm]
far beyond the supergravity limit
}

\vspace*{15mm}

{\Large Lorenzo De Lillo${}^{\,a}$ and Paolo Vallarino${}^{\,b}$}

\vspace*{8mm}
	
${}^a$ Dipartimento di Fisica, Universit\`a di Torino and INFN Sezione di Torino, \\
			Via P. Giuria 1, I-10125 Torino, Italy
			\vskip 0.3cm
			
${}^b$  Theoretische Natuurkunde, Vrije Universiteit Brussel and The International Solvay Institutes,\\ 
            Pleinlaan 2, B-1050 Brussels, Belgium

\vskip 0.8cm
	{\small
		E-mail:
		\texttt{lorenzo.delillo@unito.it;paolo.vallarino@vub.be}
	}
\vspace*{0.8cm}
\end{center}

\begin{abstract}

We study the Bremsstrahlung function in two four-dimensional $\mathcal{N}=2$ superconformal gauge theories: the \textbf{E}-theory, with hypermultiplets in the symmetric and antisymmetric representations, and the \textbf{D}-theory, with antisymmetric and fundamental matter. Using matrix-model techniques, we derive exact results at arbitrary coupling for the first few orders in the large-$N$ expansion. At leading order, both theories are shown to be planar equivalent to $\mathcal{N}=4$ Super Yang-Mills, while beyond the planar limit their behavior differs significantly. For the \textbf{E}-theory, we prove that the subleading correction is exactly given by a derivative of the free energy with respect to the coupling. This relation enables us to determine the complete strong-coupling expansion for the Bremsstrahlung function, including non-perturbative contributions, by exploiting known results for the free energy. For the \textbf{D}-theory, the exact expressions have a more intricate structure. We perform a detailed analysis of the strong-coupling regime of the subleading large-$N$ corrections, showing that, in this case, there are expansions in inverse powers of the coupling which truncate after finitely many terms. We further analyze the non-perturbative effects at strong coupling and derive closed-form expressions also for these contributions.

These results provide precise predictions for the holographic duals of both theories beyond the supergravity approximation, probing higher orders in the large-$N$ expansion and the full subleading corrections at strong coupling.
        
\end{abstract}
\end{titlepage}
\setcounter{tocdepth}{2}
\tableofcontents

\section{Introduction}
\label{sec:intro}

In quantum field theories (QFTs) the Bremsstrahlung function $B$ is defined as the coefficient governing the radiation emitted by an accelerated probe. It extends the familiar notion from Maxwell's theory of energy loss by an accelerated charged particle, as described by the Larmor and Liénard formulas. In gauge theories, however, $B$ is of particular interest because it also emerges in the analysis of $\varphi$-cusped Wilson loops $W_\varphi$, where it appears as the leading term in the small-angle expansion of the cusp anomalous dimension $\Gamma(g,\varphi)$ \cite{Polyakov:1980ca}:
\begin{align}
\big\langle W_\varphi \big\rangle \sim \rme^{-\Gamma(g,\varphi)\log\frac{\mu}{\Lambda}} \,, \qquad\quad\Gamma\left(g,\varphi\right) \underset{\varphi \to 0}{\sim} -B(g)\,\varphi^2 + O\left(\varphi^4 \right)\,,
\end{align}
where $g$ is the gauge coupling, and $\mu$ and $\Lambda$ are UV and IR cut-off scales, respectively. Deriving an exact formula for the Bremsstrahlung function in a general QFT is a highly challenging task. Nonetheless, remarkable progress has been made in superconformal field theories (SCFTs), that is, in QFTs enjoying both supersymmetry and conformal invariance. In the maximally supersymmetric case in four dimensions, $\mathcal{N}=4$ Super Yang-Mills (SYM) theory, an exact formula for $B$ was obtained in \cite{Correa:2012at,Fiol:2012sg}
\begin{align}
B_{\mathcal{N}=4} \,=\, \frac{1}{2\pi^2}\,\lambda\,\partial_\lambda \log \big\langle W \big\rangle \,,
\label{BN=4}
\end{align}
where $\lambda \equiv g^2 N$ denotes the 't~Hooft coupling. This elegant equation expresses the Bremsstrahlung function entirely in terms of a derivative with respect to the coupling of the expectation value of the $\frac{1}{2}$-BPS circular Wilson loop $W$. Its importance lies in the fact that the latter can be computed by means of supersymmetric localization, thus yielding exact results at arbitrary coupling. This finding is consistent with earlier achievements within the AdS/CFT correspondence \cite{Mikhailov:2003er,Athanasiou:2010pv,Hatta:2011gh,Fiol:2011zg} and was subsequently confirmed by integrability methods \cite{Correa:2012hh,Drukker:2012de,Gromov:2012eu}. Interestingly, the Bremsstrahlung function in $\mathcal{N}=4$ SYM\footnote{Exact results have also been obtained in the maximally supersymmetric three-dimensional theory, namely ABJM theory. Important developments can be found in \cite{Lewkowycz:2013laa,Drukker:2009hy,Forini:2012bb,Aguilera-Damia:2014qgy,Bianchi:2014laa,Bianchi:2017ozk,Bianchi:2017svd,Bianchi:2018scb}.} has been shown to be related to other physical observables. In \cite{Correa:2012at} it was proved that it determines the normalization of the two-point function of the displacement operator $\mathbb{D}_i$ on the Wilson loop as
\begin{align}
\big\langle \mathbb{D}_{i}(\tau)\,\mathbb{D}_{j}(0)\big\rangle_W\,=\,\frac{12\,B_{\mathcal{N}=4}}{\tau^4}\, \delta_{i j} \,,
\end{align}
where $i, j$ label the directions transverse to the defect. Furthermore, as shown in \cite{Lewkowycz:2013laa}, $B_{\mathcal{N}=4}$ is directly linked to the one-point function of the stress-energy tensor in the presence of the Wilson line through the relation
\begin{align}
\big\langle T_{00}(x)\big\rangle_W=\frac{B_{\mathcal{N}=4}}{3\,|x_\perp|^4} \,,
\end{align}
where $x_\perp$ denotes the distance orthogonal to the Wilson line. These important results for the Bremsstrahlung function have also been extended to theories with a lower degree of supersymmetry, in particular to four-dimensional $\mathcal{N}=2$ SCFTs. Remarkably, an exact formula for computing $B$ in this class of models was conjectured in \cite{Fiol:2015spa} and later proven in \cite{Bianchi:2018zpb,Bianchi:2019dlw} (see \cite{Galvagno:2021qyq} for a comprehensive review). The result reads
\begin{align}
B_{\mathcal{N}=2} \,=\, \frac{1}{4\pi^2}\,\partial_b \log\langle W_b\rangle\big\vert_{b=1}\,.
\label{BN=2}
\end{align}
This expression relates the Bremsstrahlung function to the vacuum expectation value of the circular Wilson loop, $W_b$, computed via localization on a squashed four-sphere. Here, $b$ denotes the squashing parameter of the ellipsoid, while the limit $b\to 1$ corresponds to the round four-sphere. Thus, also in this case, the Bremsstrahlung function can be evaluated through matrix model techniques arising from localization. 

In $\mathcal{N}=2$ theories, perturbative results have been obtained in \cite{Fiol:2015spa,Mitev:2015oty,Gomez:2018usu,Bianchi:2019dlw}. In addition, \cite{Mitev:2015oty} made a first attempt to study the strong-coupling regime of a specific $\mathcal{N}=2$ quiver gauge theory through an effective coupling approach, while \cite{Fiol:2015mrp} investigated the supergravity regime, i.e. leading large-$N$ and large-$\lambda$ limit, of $B$ in some $\mathcal{N}=2$ SCFTs. Nevertheless, explicit expressions for the Bremsstrahlung as an exact function of the coupling, as well as a corresponding precise analysis of the strong-coupling regime, are still missing. Filling this gap is particularly important also from the AdS/CFT perspective: in theories with known holographic duals, such results would provide access to a wide range of string theory effects that are currently beyond the reach of direct methods.

With this motivation, in this paper we derive expressions for $B$ that are valid for all values of the 't Hooft coupling, at the first few orders in the planar expansion, in two different $\mathcal{N}=2$ theories. Furthermore, exploiting analytical techniques, we carry out a detailed and complete study of its strong-coupling behavior, including both perturbative and non-perturbative contributions. Besides yielding explicit strong-coupling results for this meaningful observable in highly non-trivial gauge theories, our analysis also provides precise predictions for the holographic dual descriptions of these models.  \\
Specifically, as a first example of an $\mathcal{N}=2$ theory, we focus on the so-called \textbf{E}-theory. This is a superconformal gauge theory with one hypermultiplet in the symmetric representation of SU($N$) and one in the antisymmetric representation. It is holographically dual to Type IIB superstring theory on AdS$_5\times S^5/\mathbb{Z}_2$ \cite{Ennes:2000fu}, and arises as a $\mathbb{Z}_2$ orbifold/orientifold projection of AdS$_5\times S^5$. Over the past few years, localization has led to many exact results for several observables in this model and in its parent $\mathcal{N}=2$ quiver gauge theory with bifundamental matter, including two- and three-point functions of chiral primary operators \cite{Rodriguez-Gomez:2016cem,Pini:2017ouj,Beccaria:2020hgy,Galvagno:2020cgq,Beccaria:2021hvt,Billo:2021rdb,Billo:2022xas,Billo:2022gmq,Billo:2022fnb,Beccaria:2022ypy,Bobev:2022grf,Billo:2022lrv} and Wilson loop correlators \cite{Fiol:2020ojn,Billo:2019fbi,Zarembo:2020tpf,Galvagno:2021bbj,Beccaria:2021ksw,Beccaria:2021vuc,Pini:2023svd}. 

Despite having only half the supersymmetry of $\mathcal{N}=4$ SYM, this theory turns out to be very close to it: we show that the leading planar limit is in fact equivalent. At the next order in the large-$N$ expansion, scaling as $O\!\left( N^{-2} \right)$ relative to the leading one, however, the distinction between the two theories becomes evident. We determine this difference exactly as a function of the 't Hooft coupling, reducing it to a $\lambda$-derivative of the \textbf{E}-theory free energy. Combining this exact result with previous studies of the strong-coupling behavior of the free energy \cite{Beccaria:2022ypy,Bajnok:2024epf,Bajnok:2024ymr,Bajnok:2024bqr,Bajnok:2025lji}, we obtain access to the full strong-coupling expansion of the Bremsstrahlung function, including both the inverse-coupling series and the non-perturbative corrections. The latter are given by asymptotic series weighted by exponentially suppressed factors of the form $\rme^{-n\,\sqrt{\lambda}}$ (with $n\in\mathbb{N}^+$) and are therefore subleading with respect to the perturbative sector in the strong-coupling regime. They are nonetheless crucial, since the expansion in inverse powers is not Borel summable, owing to the presence of singularities on the positive real axis of the Borel plane. Their inclusion is thus necessary in order to fully reconstruct the analytic strong-coupling solution. Furthermore, these contributions acquire a natural physical interpretation as worldsheet instantons from the holographic perspective. 

As a second $\mathcal{N}=2$ model, we consider the \textbf{D}-theory, whose matter content comprises two hypermultiplets in the antisymmetric representation of SU($N$) and four in the fundamental one. In Type IIB superstring theory this model can be engineered with $N$ D3-branes in a $\mathbb{Z}_2$-orbifold probing an O7-orientifold background with four D7-branes plus their four orientifold images. The main difference compared to the \textbf{E}-theory is that the holographic dual of this $\mathcal{N}=2$ theory also includes a “D7-sector”, consisting of open string states propagating on the world-volume of the eight D7-branes with geometry of AdS$_5 \times S^3$. In this set-up, previous localization results for different observables can be found in\footnote{In \cite{Billo:2026lzb} localization computations were extended to non-conformal deformations of the \textbf{D}-theory, where the number of fundamental flavors lies in the range $0 \leq N_F < 4$.} \cite{Beccaria:2021ism,Billo:2024ftq,DeLillo:2025stg,Billo:2026lzb}. \\
We prove that, in this theory as well, the Bremsstrahlung function is planar equivalent to that of $\mathcal{N}=4$ SYM. Beyond the planar limit, however, the structure of the subleading contributions becomes significantly more intricate. First, the term that is suppressed by $O\!\left(N^{-1}\right)$ with respect to the leading large-$N$ contribution is no longer vanishing. In addition, the correction of order $O\!\left(N^{-2}\right)$ consists of the same contribution found in the \textbf{E}-theory, supplemented by further non-trivial terms. In spite of this increased complexity, we are still able to obtain exact expressions in $\lambda$. We then investigate the strong-coupling behavior of these new contributions and find that their expansions terminate after a finite number of terms, or even vanish. It then becomes important to evaluate explicitly the exponentially suppressed effects in this case as well. While a possible way to extract non-perturbative contributions for finite series of this kind is provided by \textit{Cheshire cat resurgence} (see, for example, \cite{Dorigoni:2017smz}), here we develop a different approach. More precisely, we show that our exact expressions can be written directly in terms of modified Bessel functions of the second kind, whose asymptotic expansions immediately capture the full non-perturbative series. This method was recently applied in \cite{DeLillo:2026gyy}, and here we generalize it to a considerably broader class of functions.

The outline of the paper is as  follows. In Section \ref{sec:loc}, we review the main features of the matrix model for a generic $\mathcal{N}=2$ gauge theory on the squashed four-sphere. We then introduce our method, based on a suitable change of basis, which enables us to derive exact expectation values even in the intricate interacting matrix models associated with the two theories of interest. We conclude the section by applying this method to the Bremsstrahlung function, obtaining exact expressions for the first three orders in the planar expansion: first in $\mathcal{N}=4$ SYM, where the known results are recovered, and then in the \textbf{E}- and \textbf{D}-theories.

In Section \ref{sec:strong}, we perform a detailed analysis of the strong-coupling regime of the exact results for both models. Since, in the case of the \textbf{E}-theory, the exact expression is directly related to the known free energy and thus immediately yields its strong-coupling behavior, our attention is mainly devoted to the \textbf{D}-theory. In this context, we explain our method for analyzing the strong-coupling regime by rewriting the exact matrix model expressions as asymptotic series in inverse powers of the coupling, supplemented by infinite sums of modified Bessel functions of the second kind, from which the exponentially suppressed corrections can be readily extracted.

Finally, in Section \ref{sec:concl}, we summarize our results and discuss possible future directions. Two appendices provide supplementary material: in Appendix \ref{app:nonpert}, we collect the explicit Bessel function expansions of all the exact expressions needed in our analysis, whereas in Appendix \ref{app:pert} we present an alternative derivation of the strong-coupling expansion in inverse powers of the coupling.

\section{Exact results via supersymmetric localization}
\label{sec:loc}

The partition function $\mathcal{Z}$ of an $\mathcal{N}=2$ SYM theory with SU($N$) gauge group on a compact space can be written, via localization, as an integral over the eigenvalues $a_u$ of a Hermitian matrix $a\in\mathfrak{su}(N)$ \cite{Pestun:2007rz}
\begin{align}
\label{Zloc}
\mathcal{Z}
=\int \prod_{u=1}^N da_u \;
\exp\!\left(-\frac{8\pi^2}{g^2}\,\tr a^2\right)\,
\big|Z^{\mathrm{1\text{-}loop}}\,Z^{\mathrm{inst}}\big|^2\,\,\delta\bigg( \sum_{u=1}^N a_u \bigg)~.
\end{align}
Here $Z^{\mathrm{1\text{-}loop}}$ accounts for the fluctuations around the localization locus, while $Z^{\mathrm{inst}}$ captures the instanton contributions \cite{Nekrasov:2002qd}. In the 't~Hooft limit, Yang-Mills instantons are exponentially suppressed, so we set $Z^{\mathrm{inst}}=1$ throughout the rest of the paper. 
The specific form of $Z^{\mathrm{1\text{-}loop}}$ depends on the matter content of the theory and on the geometry of the space.

For the squashed four-sphere with squashing parameter $b$\,\footnote{The squashed sphere is characterized by two radii, $\ell=bR$ and $\tilde{\ell}=R/b$, so that $b=\sqrt{\ell/\tilde{\ell}}$ and $\ell\,\tilde{\ell}=R^2$. Throughout this work we set the overall scale $R=1$.}, the one-loop determinant is \cite{Hama:2012bg}
\begin{align}
\label{Z1loopIN2*}
\big|Z_b^{\mathrm{1\text{-}loop}}\big|^2
= \frac{\displaystyle \prod_{\alpha\in\Delta}\Upsilon_b\big(\ii\,a\!\cdot\!\alpha\big)}
{\displaystyle \prod_{I}\prod_{\rho\in\mathcal{R}_I}
\Upsilon_b\!\left(\frac{Q}{2}+\ii\,a\!\cdot\!\rho\right)}\,,
\end{align}
where the product in the numerator runs over the all the roots $\alpha\in\Delta$ of the gauge Lie algebra, while the product in the denominator runs over the weights $\rho$ of the representation $\mathcal{R}_I$ carried by the $I$-th hypermultiplet. Here
\begin{align}
Q \equiv b+\frac{1}{b}\,,
\end{align}
and the special function $\Upsilon_b(x)$, originally introduced in a different context in \cite{Zamolodchikov:1995aa}, is related to the Barnes double Gamma functions and admits the integral representation \cite{Nakayama:2004vk}
\begin{align}
\label{integral}
\log \Upsilon_b(x)
=\int_{0}^{\infty}\!\frac{d\omega}{\omega}\left[
\rme^{-2\omega}\!\left(\frac{Q}{2}-x\right)^{\!2}
-\frac{\sinh^{2}\!\left(\omega\left(\frac{Q}{2}-x\right)\right)}
{\sinh(b\omega)\,\sinh(\omega/b)}\right] .
\end{align}
In the limit $b\to 1$, the one-loop determinant reduces to the $S^4$ expression,
\begin{align}
\label{Z1loop0}
\big|Z_b^{\mathrm{1\text{-}loop}}\big|^2\Big|_{b\to 1}
= \Delta(a)\,\big|Z_{S^4}^{\mathrm{1\text{-}loop}}\big|^2 = \Delta(a)\,
\frac{\displaystyle \prod_{\alpha\in\Delta} H\big(\ii\,a\!\cdot\!\alpha\big)}
{\displaystyle \prod_I \prod_{\rho\in\mathcal{R}_I} H\big(\ii\,a\!\cdot\!\rho\big)}\,,
\end{align}
where $\Delta(a)$ is the Vandermonde determinant and
\begin{align}
H(x)=\rme^{(1+\gamma)x^{2}}\,G(1+x)\,G(1-x)\,,
\end{align}
with $\gamma$ the Euler--Mascheroni constant and $G$ the Barnes $G$-function. In particular, for $\mathcal{N}=4$ SYM the one-loop determinant simplifies to
$\big|Z_b^{\mathrm{1\text{-}loop}}\big|^2\big|_{b\to 1}=\Delta(a)$, so that the matrix model \eqref{Zloc} becomes purely Gaussian\,\footnote{In \cite{DeLillo:2025stg}, the one-loop determinant for $\mathcal{N}=2$ theories on the squashed sphere was rewritten in terms of functions $H_v(x;b)$ and $H_h(x;b)$, associated with vector and hypermultiplets, respectively. Compared to $\Upsilon_b(x)$, these functions exhibit properties and admit expansions that more closely parallel those of the Barnes $G$-function.}.

Furthermore, the special function $\Upsilon_b(x)$ obeys the relation
\begin{align}
\Upsilon_b(x)=\Upsilon_{1/b}(x)\,,
\end{align}
which renders the expression in \eqref{Z1loopIN2*} invariant under the exchange $b\leftrightarrow 1/b$. This symmetry furnishes important constraints on the partition function: expanding around the round-sphere point $b=1$ immediately shows that the linear term must vanish, i.e.
\begin{align}
\partial_b \mathcal{Z}_b\big|_{b=1}=0\,,
\label{firstder}
\end{align}
and the first non-trivial corrections arise at second order in $(b-1)$, as explicitly shown in \cite{DeLillo:2025stg}.

\subsection{Bremsstrahlung function}
Let us now discuss how to evaluate the Bremsstrahlung function \eqref{BN=2} within the localized matrix model of a generic $\mathcal{N}=2$ SYM theory. On the ellipsoid there are two distinct $\tfrac12$-BPS circular Wilson loops in the fundamental representation: one wrapping the circle of radius $b$ in the $(x^1,x^2)$ plane, while the other one wrapping the circle of radius $1/b$ in the $(x^3,x^4)$ plane. In \cite{Hama:2012bg} it was shown that these operators localize to the matrix model as
\begin{align}
W_b \,=\, \frac{1}{N}\tr \exp\!\left(-2\pi b\, a\right)\,,\qquad 
W_{1/b} \,=\, \frac{1}{N}\tr \exp\!\left(-2\pi b^{-1} a\right)\,,
\label{Wb}
\end{align}
and both reduce to the standard circular Wilson loop on the round sphere, $W$, when $b=1$. Their expectation value is therefore
\begin{align}
\big\langle W_b \big\rangle
= \frac{1}{\mathcal{Z}_b}\int \prod_{u=1}^N da_u \; W_b \;
\exp\!\left(-\frac{8\pi^2}{g^2}\,\tr a^2\right)\,
\big|Z_{b}^{\mathrm{1\text{-}loop}}\big|^2\,\delta\bigg( \sum_{u=1}^N a_u \bigg) \, ,
\end{align}
and similarly for $W_{1/b}$.

We can now compute the Bremsstrahlung function from \eqref{BN=2}:
\begin{align}
B_{\mathcal{N}=2}
&
= \frac{1}{4\pi^2}\,
\frac{\partial_b\big\langle W_b\big\rangle\big|_{b=1}}{\big\langle W\big\rangle}
= \frac{1}{4\pi^2}\,
\frac{\big\langle \partial_b W_b\big|_{b=1}\big\rangle}{\big\langle W\big\rangle}\,\equiv\,\frac{1}{4\pi^2}\,\frac{\big\langle W' \big\rangle}{\big\langle W\big\rangle}\,,
\label{BN=2new}
\end{align}
where in the second equality we used \eqref{firstder} to drop the contribution from $\partial_b \mathcal{Z}_b$ at $b=1$. This simplification is crucial: as already observed in \cite{Fiol:2015spa}, it implies that $B$ can be extracted using only the $S^4$ matrix model, i.e.\ the one-loop determinant evaluated at $b=1$. For this reason, we obtain
\begin{align}
\big\langle W' \big\rangle\,
= \,\frac{1}{\mathcal{Z}_{S^4}}\int \prod_{u=1}^N da_u \,\Delta(a)\, W' \,
\exp\!\left(-\frac{8\pi^2}{g^2}\,\tr a^2\right)\,
\frac{\prod_{\alpha\in\Delta} H\!\big(\ii\,a\!\cdot\!\alpha\big)}
{\prod_I \prod_{\rho\in\mathcal{R}_I} H\!\big(\ii\,a\!\cdot\!\rho\big)}\,\delta\bigg( \sum_{u=1}^N a_u \bigg)\, .
\label{wmm}
\end{align}
We now perform the rescaling\,\footnote{We included an overall minus sign in the rescaling so that the Wilson loop operator appears with a positive exponent. This choice is purely conventional.}
\begin{align}
a \;\longrightarrow\; -\sqrt{\frac{\lambda}{8\pi^2 N}}\;a \, ,
\label{rescaling}
\end{align}
and adopt the "full Lie algebra" approach \cite{Billo:2017glv}. With this choice, the partition function of $\mathcal{N}=2$ SYM theories on the four-sphere can be written, up to an overall normalization constant, as\,\footnote{We decompose $a=a^cT_c$ with $c=1,\ldots,N^2-1$, where $T_c$ are the $\mathrm{SU}(N)$ generators in the fundamental representation, normalized as $\tr(T_bT_c)=\tfrac12\delta_{bc}$. The integration measure is
\begin{align*}
da=\prod_{c=1}^{N^2-1}\frac{da^c}{\sqrt{2\pi}}
\qquad\text{so that}\qquad
\int da\,\rme^{-\tr a^2}=1\,.
\end{align*}}
\begin{align}
\mathcal{Z}_{S^4}
=\int da\;\rme^{-\tr a^2 - S_{\rm int}}\,,
\label{ZS}
\end{align}
where $S_{\rm int}$ encodes the interaction terms of the matrix model. It vanishes identically for $\mathcal{N}=4$ SYM, while for $\mathcal{N}=2$ theories it depends on the matter content. In the cases relevant for us it takes the form \cite{Beccaria:2020hgy}
\begin{align}
S^{\E}
&=4\sum_{n=1}^{\infty}\sum_{k=1}^{n-1}(-1)^{n}\,
\frac{(2n+1)!\,\zeta(2n+1)}{(2k+1)!\,(2n-2k+1)!}\,
\Big(\frac{\lambda}{8\pi^2N}\Big)^{n+1}\,
\tr a^{2k+1}\, \tr a^{2n-2k+1}\,,
\label{Se}\\[2mm]
S^{\D}
&=S^{\E}-4\sum_{n=1}^{\infty} (-1)^{n}
\frac{(4^{n}-1)\,\zeta(2n+1)}{n+1}\,
\Big(\frac{\lambda}{8\pi^2N}\Big)^{n+1}\,
\tr a^{2n+2}\,.
\label{Sd}
\end{align}
The expectation value of a generic operator $\mathcal{O}$ in the interacting matrix model can then be expressed as
\begin{align}
\big\langle \mathcal{O}\big\rangle
=\frac{\big\langle \rme^{-S_{\rm int}}\,\mathcal{O}\big\rangle_{0}}
{\big\langle \rme^{-S_{\rm int}}\big\rangle_{0}}\,,
\label{vev}
\end{align}
where $\langle\cdot\rangle_{0}$ denotes expectation values in the Gaussian matrix model.

In particular, we are interested in the circular Wilson loop in the fundamental representation $W_b$. Using its definition \eqref{Wb} together with the rescaling \eqref{rescaling}, one finds
\begin{align}
\label{defWL}
W_b
=\frac{1}{N}\sum_{k=0}^{\infty}\frac{b^k}{k!}
\left(\frac{\lambda}{2N}\right)^{\frac{k}{2}}\tr a^{k}\,,
\end{align}
which reduces, for $b=1$, to the standard circular Wilson loop $W$ on $S^4$ \cite{Pestun:2007rz}. Differentiating with respect to $b$ and then setting $b=1$ gives the corresponding insertion
\begin{align}
\label{defWLprime}
W' \equiv \partial_b W_b\big|_{b=1}
= \frac{1}{N}\sqrt{\frac{\lambda}{2N}}\,
\tr\!\left(a\,\exp\!\left[\sqrt{\frac{\lambda}{2N}}\,a\right]\right)
=\frac{1}{N}\sum_{k=1}^{\infty}\frac{1}{(k-1)!}
\left(\frac{\lambda}{2N}\right)^{\frac{k}{2}}\tr a^{k}\, .
\end{align}
For later convenience, it is also useful to note the identity
\begin{align}
W' \,=\, 2\lambda\,\partial_\lambda W\,.
\label{idWL}
\end{align}

Having established the required ingredients, we proceed to compute the Bremsstrahlung function \eqref{BN=2new} in the matrix model, and in particular to analyze its large-$N$ limit. A useful way to streamline the planar expansion is to perform a change of basis. Thus we introduce a new set of operators $\mathcal{P}_k$ as follows \cite{Rodriguez-Gomez:2016cem,Beccaria:2020hgy}
\begin{align}
\label{changebasis}
\tr a^k
=\Big(\frac{N}{2}\Big)^{\frac{k}{2}}
\sum_{\ell=0}^{\lfloor\frac{k-1}{2}\rfloor}
\sqrt{k-2\ell}\,
\frac{k!}{\ell!\,(k-\ell)!}\,
\mathcal{P}_{k-2\ell}
\;+\;
\big\langle \tr a^k \big\rangle_0 \, ,
\end{align}
where the Gaussian expectation values read
\begin{align}
\big\langle \tr a^{2n+1} \big\rangle_0 &= 0~, \notag \\
\big\langle \tr a^{2n} \big\rangle_0
&= \frac{N^{n+1}}{2^n}\,\frac{(2n)!}{n!\,(n+1)!}
+\frac{N^{n-1}}{2^{n+1}}\,\frac{(2n)!}{n!\,(n-1)!}\,
\Big(\frac{n-1}{6}-1\Big)
+O\!\big(N^{n-3}\big)~.
\label{tracesVEV}
\end{align}
The operators $\mathcal{P}_k$ enjoy a number of useful properties in the Gaussian matrix model. By construction they are normal-ordered, so that their one-point functions vanish identically,
\begin{align}
\big\langle \mathcal{P}_k \big\rangle_0 = 0\,,
\label{1ptN=4}
\end{align}
and they are particularly well suited to the planar limit: they form an orthonormal basis up to $1/N^2$ corrections,
\begin{align}
\big\langle \mathcal{P}_k\,\mathcal{P}_\ell\big\rangle_0\,=\,\delta_{k,\ell}+O\big(N^{-2}\big)\,.
\end{align}
This basis also provides an efficient organization of the large-$N$ expansion in interacting matrix models. Implementing the change of basis in a generic $\mathcal{N}=2$ SYM theory, the relevant operators for us take the form
\begin{align}
\big\langle W\big\rangle\,=\,& \big\langle W \big\rangle_0\,+\,\frac{1}{N}\sum_{n=2}^\infty \sqrt{n}\,I_n( \sqrt{\lambda})\,\big\langle \mathcal{P}_n \big\rangle \,, \label{WL} \\
\big\langle W'\big\rangle \,=\,& \big\langle W' \big\rangle_0\,+\, \frac{2\,\lambda}{N}\sum_{n=2}^\infty \sqrt{n}\,\left(\partial_{\lambda}I_n( \sqrt{\lambda} )\right)\big\langle \mathcal{P}_n \big\rangle \,, \label{WL'}
\end{align}
where $I_n(x)$ is a modified Bessel function of the first kind. Therefore, from these formulae it is straightforward to understand that in any $\mathcal{N}=2$ theory the Bremsstrahlung function is completely fixed by the one-point function of the operator $\mathcal{P}_k$,  computed in the interacting matrix model of the given theory using \eqref{vev}.

After briefly reviewing the $\mathcal{N}=4$ SYM case, we are going to analyze the Bremsstrahlung function in the \textbf{E}- and \textbf{D}-theories. For each model, we determine this quantity to the first orders in the planar limit. As in $\mathcal{N}=4$ \cite{Correa:2012at}, we obtain an exact expression in the ’t~Hooft coupling by resumming the perturbative series, and we then extract the corresponding strong-coupling expansions.

Specifically, in all the theories we will consider, the large-$N$ expansions take the form
\begin{align}
\big\langle W \big\rangle\,=\,W^{(0)}+\frac{W^{(1)}}{N}+\frac{W^{(2)}}{N^2}+O\!\left(N^{-3}\right)\,,\quad \big\langle W' \big\rangle\,=\,W'^{(0)}+\frac{W'^{(1)}}{N}+\frac{W'^{(2)}}{N^2}+O\!\left(N^{-3}\right)\,,
\label{expW}
\end{align}
where all coefficients are exact functions of the 't~Hooft coupling.
Using \eqref{BN=2new}, we then obtain
\begin{align}
B
\,=\,B^{(0)}+\frac{B^{(1)}}{N}+\frac{B^{(2)}}{N^2}+O\!\left(N^{-3}\right)\,,
\label{expBseries}
\end{align}
with
\begin{align}
B^{(0)} &= \frac{1}{4\,\pi^2}\,\frac{W'^{(0)}}{W^{(0)}}\,, \qquad B^{(1)} = \frac{1}{4\,\pi^2}\,\frac{W'^{(1)}\,W^{(0)} - W'^{(0)}\,W^{(1)}}{\left(W^{(0)}\right)^2} ,\\[4pt]
B^{(2)} &= \frac{1}{4\,\pi^2}\,\frac{ W'^{(2)}\left(W^{(0)}\right)^2
          - W'^{(1)}\,W^{(0)}\,W^{(1)}
          + W'^{(0)}\left(W^{(1)}\right)^2
          - W'^{(0)}\,W^{(0)}\,W^{(2)}}
          {\left(W^{(0)}\right)^3}\,.
\label{expB}
\end{align}
In general, the coefficients $B^{(n)}$ obey the recursion relation
\begin{align}
    B^{(n)}= \frac{1}{4\pi^2} \frac{W'^{(n)}}{W^{(0)}} - \sum_{i=1}^n \frac{W^{(i)}}{W^{(0)}} B^{(n-i)} \,.
\end{align}

\subsection{\texorpdfstring{$\mathcal{N}=4$}{} SYM}

As a first check, one can verify that the $\mathcal{N}=2$ relation \eqref{BN=2}
reduces to \eqref{BN=4} upon specialization to $\mathcal{N}=4$.
This is particularly transparent in the matrix model description:
\begin{align}
B_{\mathcal{N}=4}
&= \frac{1}{4\pi^2}\,\partial_b \log\big\langle W_b\big\rangle_0\big|_{b=1}
= \frac{1}{4\pi^2}\,\frac{\big\langle W' \big\rangle_0}{\langle W\rangle_0}
= \frac{\lambda}{2\pi^2}\,\frac{\big\langle \partial_\lambda W \big\rangle_0}{\langle W\rangle_0}
= \frac{1}{2\pi^2}\,\lambda\,\partial_\lambda \log\big\langle W \big\rangle_0 \, .
\end{align}
In the first equality we used \eqref{BN=2new}, in the second the identity \eqref{idWL},
and in the last step we exploited the fact that, for $\mathcal{N}=4$ SYM, the matrix model is Gaussian,
so the coupling dependence resides solely in the Wilson loop operator.

The expectation value of the $\tfrac12$-BPS circular Wilson loop is known exactly as a function of $\lambda$ and $N$
\cite{Drukker:2000rr,Erickson:2000af}. Using \eqref{BN=4}, the corresponding Bremsstrahlung function was obtained in
\cite{Correa:2012at}. Now let us recall the first orders in the large-$N$ expansion of $\big\langle W \big\rangle_0$:
\begin{align}
W^{(0)}_{\mathcal{N}=4}
&= \frac{2}{\sqrt{\lambda}}\,I_1(\sqrt{\lambda})\,,
\qquad
W^{(1)}_{\mathcal{N}=4}=0\,,
\label{WLON=4}\\[0.5em]
W^{(2)}_{\mathcal{N}=4}
&= \frac{1}{48}\!\left(\lambda\,I_0(\sqrt{\lambda})
-14\sqrt{\lambda}\,I_1(\sqrt{\lambda})\right)\,.
\label{WNLON=4}
\end{align}
These expressions can be immediately recovered by inserting the Gaussian expectation values \eqref{tracesVEV}  in the definition \eqref{defWL} at $b=1$. In the same way one finds 
\begin{align}
W'^{(0)}_{\mathcal{N}=4}
&= 2\,I_2(\sqrt{\lambda})\,,
\qquad
W'^{(1)}_{\mathcal{N}=4}=0\,,
\label{W'LON=4}\\[0.5em]
W'^{(2)}_{\mathcal{N}=4}
&= \frac{1}{48}\!\left(-12\lambda\,I_0(\sqrt{\lambda})
+\lambda^{\frac{3}{2}}\,I_1(\sqrt{\lambda})\right)\,.
\label{W'NLON=4}
\end{align}
It then follows that\,\footnote{A useful identity for modified Bessel functions of the first kind is $I_0(x)\,=\,I_2(x)+\frac{2}{x}I_1(x)$.}
\begin{align}
B_{\mathcal{N}=4}^{(0)}
&= \frac{\sqrt{\lambda}}{4\pi^2}\,\frac{I_2(\sqrt{\lambda})}{I_1(\sqrt{\lambda})}\,,
\qquad
B_{\mathcal{N}=4}^{(1)}=0\,,
\label{BLON=4}\\[0.5em]
B_{\mathcal{N}=4}^{(2)}
&= \frac{\lambda}{384\,\pi^2}\!\left( \lambda-24-\lambda\,\frac{I_2(\sqrt{\lambda})^2}{I_1(\sqrt{\lambda})^2} \right)\,.
\label{BNLON=4}
\end{align}
This reproduces the known result of \cite{Correa:2012at}. We now move on to $\mathcal{N}=2$ gauge theories, for which the matrix model is genuinely interacting: the potential is highly non-trivial, and obtaining closed-form expressions in the ’t~Hooft coupling becomes much more involved. Nonetheless, we will show that for our two specific models, by exploiting the decompositions in \eqref{WL} and \eqref{WL'}, it is still possible to resum the perturbative series and derive exact expressions in $\lambda$ for the first orders in the large-$N$ expansion.

\subsection{\texorpdfstring{$\mathbf{E}$}{}-theory}

The matrix model interaction action of the \textbf{E}-theory is given in \eqref{Se}. In order to compute the Bremsstrahlung function, it is convenient to note from \eqref{WL}--\eqref{WL'} that the theory-dependent input enters through the one-point functions of the operators $\mathcal{P}_k$. These expectation values can be evaluated in the full Lie-algebra formalism starting from \eqref{vev}. The result is \cite{Billo:2024ftq}
\begin{align}
\big\langle \mathcal{P}_{2k} \big\rangle_{\mathbf{E}}
&=\,-\frac{\sqrt{2k}}{N}\,\lambda\,\partial_\lambda \mathcal{F}_{\mathbf{E}}
+O\!\left(N^{-3}\right)\,,
\qquad
\big\langle \mathcal{P}_{2k+1} \big\rangle_{\mathbf{E}}=0\,,
\label{1ptE}
\end{align}
where $\mathcal{F}_{\mathbf{E}}$ denotes the free energy of the \textbf{E}-theory. To leading order in the large-$N$ expansion it can be written in closed form as \cite{Beccaria:2021hvt}
\begin{align}
\mathcal{F}_{\mathbf{E}}
=\frac{1}{2}\,\Tr \log \big(1-\mathsf{X}\big)
+O\!\left(N^{-2}\right)\,,
\label{freeenergyE}
\end{align}
with $\mathsf{X}$ an infinite matrix whose entries are given by a convolution of Bessel functions
\begin{align}
\mathsf{X}_{k,\ell}
= 8\,(-1)^{k+\ell+1}\sqrt{(2k+1)(2\ell+1)}
\int_{0}^{\infty}\!\frac{dt}{t}\,
\frac{\rme^{t}}{(\rme^{t}-1)^2}\,
J_{2k+1}\!\Big(\frac{t\sqrt{\lambda}}{2\pi}\Big)\,
J_{2\ell+1}\!\Big(\frac{t\sqrt{\lambda}}{2\pi}\Big)\,,
\label{Xmatrix}
\end{align}
for $k,\ell\,\geq\,1$. 

Combining \eqref{1ptE} with \eqref{WL}, one obtains the first coefficients in the large-$N$ expansion of the Wilson loop \footnote{To derive \eqref{WNLOE} we used the Bessel function identity
\begin{align}
\sum_{n=1}^\infty (2n)\, I_{2n}(x)\,=\,\frac{x}{2}\,I_1(x)\,.
\label{sumid}
\end{align}}
\begin{align}
W^{(0)}_{\mathbf{E}}
&= W^{(0)}_{\mathcal{N}=4}\,,
\qquad
W^{(1)}_{\mathbf{E}}=W^{(1)}_{\mathcal{N}=4}=0\,,
\label{WLOE}\\[0.5em]
W^{(2)}_{\mathbf{E}}
&= W^{(2)}_{\mathcal{N}=4}
-\frac{\lambda^{\frac{3}{2}}}{2}\,I_1(\sqrt{\lambda})\,\partial_\lambda\mathcal{F}_\mathbf{E}\,.
\label{WNLOE}
\end{align}
In the planar limit, the \textbf{E}-theory matches $\mathcal{N}=4$ for this observable, while deviations between the two theories first appear at order $O(N^{-2})$. The same result was previously obtained in \cite{Beccaria:2021vuc} by different methods. It is worth stressing that \eqref{WNLOE} is exact in the 't~Hooft coupling: the genuinely $\mathcal{N}=2$ contribution is entirely captured by the term proportional to $\partial_\lambda\mathcal{F}_\mathbf{E}$.

Let us now evaluate the new contribution, i.e. the $b$-derivative Wilson loop expectation value, which is needed to determine $B$. Using \eqref{WL'} one finds
\begin{align}
W'^{(0)}_{\mathbf{E}}
&= W'^{(0)}_{\mathcal{N}=4}\,,
\qquad
W'^{(1)}_{\mathbf{E}}=W'^{(1)}_{\mathcal{N}=4}=0\,,
\label{W'LOE}\\[0.5em]
W'^{(2)}_{\mathbf{E}}
&= W'^{(2)}_{\mathcal{N}=4}  -\frac{\lambda^2}{2}\,I_0(\sqrt{\lambda})\,\partial_\lambda\mathcal{F}_\mathbf{E} \,.
\label{W'NLOE}
\end{align}
Combining these results with the expansion \eqref{expB}, we end up with
\begin{align}
& B_{\mathbf{E}}^{(0)}=B_{\mathcal{N}=4}^{(0)}\,,
\qquad
B_{\mathbf{E}}^{(1)}=B_{\mathcal{N}=4}^{(1)}=0\,,
\label{BLOE}\\[0.5em]
& B_{\mathbf{E}}^{(2)}= B_{\mathcal{N}=4}^{(2)} - \frac{\lambda^2}{8\,\pi^2}\,\partial_\lambda\mathcal{F}_\mathbf{E} \,.
\label{BNLOE}
\end{align}
This provides the final expression for the Bremsstrahlung function in the \textbf{E}-theory. Relative to $\mathcal{N}=4$ SYM, the entire $\mathcal{N}=2$ correction is encoded in the $\lambda$-derivative of the free energy, which is known exactly in terms of a Bessel function convolution \eqref{freeenergyE}. This expression gives us full analytic control in both the perturbative regime and the strong-coupling limit.

\subsection{\texorpdfstring{$\mathbf{D}$}{}-theory}

In this case the interaction action in the matrix model is provided in \eqref{Sd}. As in the \textbf{E}-theory, the one-point function of the operator $\mathcal{P}_k$ can be computed from \eqref{vev}. One finds \cite{Billo:2024ftq}
\begin{align}
\big\langle \mathcal{P}_{2k} \big\rangle_{\mathbf{D}}
&=\,\mathsf{Y}_{2k}
+\frac{\sqrt{2k}}{2\,N}\,\mathsf{Y}^2
-\frac{\sqrt{2k}}{N}\,\lambda\,\partial_\lambda \mathcal{F}_{\mathbf{E}}
+O\!\left(N^{-2}\right)\,,
\qquad
\big\langle \mathcal{P}_{2k+1} \big\rangle_{\mathbf{D}}=0\,,
\label{1ptD}
\end{align}
where $\mathsf{Y}_{2k}$ admits the integral representation
\begin{align}
\label{Yvector}
\mathsf{Y}_{2k}
=\, &(-1)^{k+1}\,2\sqrt{2k}\int_0^\infty \!\mathcal{D}t\,
J_{2k}\Big(\frac{\sqrt{\lambda}\,t}{\pi}\Big)
-\delta_{k,1}\,\frac{\sqrt{2}\log 2}{4\pi^2}\,\lambda\,,
\\[0.5em]
\mathsf{Y}
\equiv\,&\sum_{k=1}^\infty \sqrt{2k}\,\mathsf{Y}_{2k}
=
\frac{\sqrt{\lambda}}{\pi}\!\int_0^\infty \!\mathcal{D}t\,t\,
J_{1}\Big(\frac{\sqrt{\lambda}\,t}{\pi}\Big)
-\frac{\log 2}{2\pi^2}\,\lambda\,,
\label{Yhat}
\end{align}
with
\begin{align}
\mathcal{D}t\,= \,&  \frac{dt}{t}\,
\frac{\rme^t}{(\rme^t+1)^2} \,. 
\label{kernel}
\end{align}
In contrast to $\mathcal{N}=4$ SYM and to the \textbf{E}-theory, the one-point function in the \textbf{D}-theory contains an $O(N^{0})$ contribution. As a result, the large-$N$ expansion of observables develops a different structure. We can thus proceed with the Wilson loop
\begin{align}
W^{(0)}_{\mathbf{D}}
&= W^{(0)}_{\mathcal{N}=4}\,, \label{WLOD} \\[0.5em]
W^{(1)}_{\mathbf{D}}&=-\frac{\lambda\,\log2}{2\pi^2}\,I_2(\sqrt{\lambda} ) + \int_0^\infty \!\mathcal{D}t\,\mathcal{Y}\left( \lambda,t \right)  \,, \label{WNLOD}\\[0.5em]
W^{(2)}_{\mathbf{D}}
&= W^{(2)}_{\mathbf{E}} + \frac{\sqrt{\lambda}}{4}\,I_1( \sqrt{\lambda})\mathsf{Y}^2 \,,
\label{WNNLOD}
\end{align}
where we have defined
\begin{align}
\mathcal{Y}\left( \lambda,t \right)\,=\,&\frac{t\,\sqrt{\lambda}\,\pi}{t^2+\pi^2}
\left(I_2\left(\sqrt{\lambda}\right)J_{1}\left(\frac{\sqrt{\lambda}\,t}{\pi}\right)+\frac{t}{\pi}I_1\left(\sqrt{\lambda}\right)J_{2}\left(\frac{\sqrt{\lambda}\,t}{\pi}\right)\right)\,. \label{Ycal}
\end{align}
In deriving \eqref{WNLOD} we used the Bessel function summation identity
\begin{align}
\sum_{n=1}^\infty (-1)^n(2n)\, I_{2n}(x)\,J_{2n}(y)
=\,-\frac{x\,y}{2(x^2+y^2)}
\left(x\,I_2(x)\,J_1(y)+y\,I_1(x)\,J_2(y) \right)\,,
\label{sumid2}
\end{align}
which follows straightforwardly from related identities (see Appendix~A of \cite{Pini:2023lyo}). This result  was previously obtained in \cite{Beccaria:2021ism} using a distinct approach. This makes explicit that, although the \textbf{D}-theory still agrees with $\mathcal{N}=4$ SYM at planar order for this observable, the first deviations already appear at $O(N^{-1})$. Crucially, these deviations are encoded in expressions that are exact in the 't~Hooft coupling~$\lambda$.
As before, we then evaluate the contribution given by the $b$-derivative term. We obtain\,\footnote{As a direct check of eqs.~\eqref{W'NLOE},
\eqref{W'NLOD}, and \eqref{W'NNLOD}, we numerically evaluated the finite-$N$ matrix model integrals for $\langle W'\rangle$ in the \textbf{E}- and \textbf{D}-theories for several finite values of $\lambda$ and
$N=2,3,4$, using \eqref{wmm}. We compared these results with the corresponding large-$N$ expressions exact in $\lambda$, truncated at order $1/N^2$. We observe that the agreement improves systematically as $N$ increases, consistently with the suppression of the higher-order terms omitted from the truncated $1/N$ expansion. We regard these numerical checks as a further validation of our large-$N$ results.}
\begin{align}
W'^{(0)}_{\mathbf{D}}
&= W'^{(0)}_{\mathcal{N}=4}\,, \label{W'LOD} \\[0.5em]
W'^{(1)}_{\mathbf{D}}&=-\frac{\lambda^2\,\log2}{\pi^2}\,\partial_\lambda I_2(\sqrt{\lambda} ) + \int_0^\infty\!\mathcal{D}t\,\left(2\lambda\partial_\lambda -\,t\,\partial_t\right)\mathcal{Y}\left( \lambda,t\right) \,, \label{W'NLOD} \\[0.5em]
W'^{(2)}_{\mathbf{D}}
&= W'^{(2)}_{\mathbf{E}} + \frac{\lambda}{4}\,I_0( \sqrt{\lambda})\mathsf{Y}^2  \,.
\label{W'NNLOD}
\end{align}
We eventually plug these results in the formula for the Bremsstrahlung \eqref{expB} and get
\begin{align}
& B_{\mathbf{D}}^{(0)}=B_{\mathcal{N}=4}^{(0)}\,, \label{BLOD} \\[1em]
& B_{\mathbf{D}}^{(1)}=\frac{1}{32\,\pi^4\,I_1(\sqrt{\lambda})^2}\biggl\{2\log 2\, \lambda^{\frac{3}{2}}\left( \sqrt{\lambda}\,I_2(\sqrt{\lambda})^2- \sqrt{\lambda}\,I_1(\sqrt{\lambda})^2+2\,I_1(\sqrt{\lambda})\,I_2(\sqrt{\lambda})  \right) \notag \\ 
&  \qquad\qquad\qquad\qquad -4\pi^2\sqrt{\lambda}\int_0^\infty\!\mathcal{D}t\,\left( \sqrt{\lambda}\,I_2(\sqrt{\lambda})\mathcal{Y}\left( \lambda,t \right)-I_1(\sqrt{\lambda}) \left( 2\lambda\partial_\lambda -t\,\partial_t\right)\mathcal{Y}\left( \lambda,t \right)\right) \biggr\}\,,
\label{BNLOD}\\[1em]
& B_{\mathbf{D}}^{(2)}= B_{\mathbf{E}}^{(2)} + \frac{\lambda}{16\,\pi^2}\,\mathsf{Y}^2 +  \frac{\lambda}{128\,\pi^6\,I_1( \sqrt{\lambda} )^3} \ \times \notag \\
& \qquad\quad \times \biggl\{8\,\pi^4 \int_0^\infty\!\mathcal{D}t\,\mathcal{Y}\left( \lambda,t \right) \int_0^\infty\!\mathcal{D}t'\biggl(\sqrt{\lambda}\,I_2( \sqrt{\lambda} )\mathcal{Y}\left( \lambda,t' \right) 
-I_1( \sqrt{\lambda} )\left( 2\lambda\partial_\lambda -t'\,\partial_t'\right)\mathcal{Y}\left( \lambda,t' \right) \biggr) \notag \\
& \qquad\qquad +2(\log 2)^2\,\lambda^{2}\,I_2(\sqrt{\lambda})\biggl( \sqrt{\lambda}\,I_2(\sqrt{\lambda})^2- \sqrt{\lambda}\,I_1(\sqrt{\lambda})^2 
+2\,I_1(\sqrt{\lambda})\,I_2(\sqrt{\lambda})   \biggr) \notag \\
& \qquad\qquad+ 4\pi^2\log 2 \,\lambda \biggl[\int_0^\infty\!\mathcal{D}t\,\mathcal{Y}\left( \lambda,t \right)\biggl(\sqrt{\lambda}\,I_1(\sqrt{\lambda})^2- 2\sqrt{\lambda}\,I_2(\sqrt{\lambda})^2 -2\,I_1(\sqrt{\lambda})\,I_2(\sqrt{\lambda})  \biggr) \notag  \\
& \qquad\qquad\qquad\qquad\qquad\qquad\qquad\qquad +I_1(\sqrt{\lambda})\,I_2(\sqrt{\lambda})  \int_0^\infty\!\mathcal{D}t\left( 2\lambda\partial_\lambda -t\,\partial_t\right)\mathcal{Y}\left( \lambda,t \right)\biggr] \biggr\} \,.
\label{BNNLOD}
\end{align}
This is the final result for the Bremsstrahlung function in the \textbf{D}-theory. Compared with the \textbf{E}-theory, which retains a close structural resemblance to $\mathcal{N}=4$ SYM, the result is more intricate. The reason is that in the \textbf{D}-theory the matrix model interaction contains, in addition to the double-trace term, a single-trace contribution reflecting the presence of fundamental hypermultiplets.

Nonetheless, the expression is again exact in the ’t~Hooft coupling~$\lambda$ and is organized in terms of elementary building blocks, consisting of integrals involving a single Bessel function. Such integrals are particularly amenable to a systematic strong-coupling analysis, as we will show in the next section.

\section{Strong-coupling analysis}
\label{sec:strong}

The matrix model techniques emerging from localization lead to explicit exact expressions for the Bremsstrahlung function, in all the models considered, expressed as integrals of Bessel functions. In particular, in the weak-coupling regime, perturbative effects are readily accessible to very high orders through the expansion of the Bessel functions around $\lambda\to 0$. In the strong-coupling regime, however, the analysis of the exact expressions becomes significantly more subtle. In this section, we present a comprehensive study of the strong-coupling behavior of the Bremsstrahlung function in the $\mathcal{N}=2$ theories under investigation.

As a starting point, it is important to recall that in $\mathcal{N}=4$ SYM theory the strong-coupling behavior is governed by the asymptotic expansion of the modified Bessel functions of the first kind, i.e.
\begin{align}
    I_\nu(z) \underset{|z| \to \infty}{\sim} \frac{e^z}{(2\pi z)^{1/2}} \sum_{j=0}^{\infty} \frac{a_j(\nu)}{z^j}\pm i e^{\pm \nu\pi i} \frac{e^{-z}}{(2\pi z)^{1/2}} \sum_{j=0}^{\infty} (-1)^j\,\frac{a_j(\nu)}{z^j}
\label{expBI}
\end{align}
with $|\text{arg}\ z|<\frac{\pi}{2}$ and
\begin{align}
 a_j(\nu) \equiv  \frac{(\frac{1}{2}+\nu)_j (\frac{1}{2}-\nu)_j}{2^j\,j!}
\label{aj}
\end{align}
where $(x)_j$ is the Pochhammer symbol. This asymptotic expansion takes the form of an inverse power series multiplied by the exponential factor $e^z$, which provides the dominant contribution. Nevertheless, this sector alone is not Borel summable, and the inclusion of the non-perturbative corrections proportional to $e^{-z}$ is necessary to recover a well-defined analytic function. A similar structure also appears in the $\mathcal{N}=2$ theories, where the strong-coupling expansion consists of a series in inverse powers of the coupling together with additional series weighted by exponentially suppressed factors. As we shall show, although the latter terms are subleading, they are crucial, since, in some cases, they contain the only non-trivial information about the strong-coupling behavior of certain functions whose perturbative asymptotic expansion vanishes identically.

We now turn to a detailed analysis of the two $\mathcal{N}=2$ theories. 

\subsubsection*{\textbf{E}-theory}

The result for the Bremsstrahlung function in the \textbf{E}-theory takes an especially compact form, since it is expressed directly in terms of a $\lambda$-derivative of the free energy \eqref{BLOE}-\eqref{BNLOE}. The free energy itself may be exactly written in terms of a semi-infinite matrix, which is a convolution of Bessel functions of the first kind, as shown in \eqref{freeenergyE}-\eqref{Xmatrix}. The strong-coupling properties of this function, including both its perturbative and non-perturbative sectors, have been thoroughly investigated in recent years. Specifically, the full perturbative strong-coupling expansion for the free energy of the $\mathbf{E}$-theory in the planar limit was found in \cite{Beccaria:2022ypy}, while the exponentially suppressed terms and their resurgence properties were studied in \cite{Bajnok:2024epf,Bajnok:2024ymr,Bajnok:2024bqr,Bajnok:2025lji}. The final result reads\,\footnote{Non-planar corrections to the free energy and to the expectation value of the circular Wilson loop have been investigated in \cite{Beccaria:2023kbl}.}
\begin{align}
\mathcal{F}_{\mathbf{E}} \underset{\lambda \to \infty}{\sim} &  \notag \frac{\sqrt{\lambda}}{8} 
- \frac{3}{8}\log \lambda +k + \frac{3}{32}\log\left(\frac{\lambda'}{\lambda}\right) 
- \frac{15\,\zeta(3)}{64 \,\lambda'^{\,3/2}} 
- \frac{945\,\zeta(5)}{512 \,\lambda'^{\,5/2}}  
+ O\left(\frac{1}{\lambda'^{\,3}}\right)  \notag \\[0.5em]
& + \frac{\ii\,\sqrt{\lambda'}}{64}\rme^{-\sqrt{\lambda'}} \left( 1+\frac{1+16\log 2}{2\,\lambda '^{\,1/2}}-\frac{15-96\log 2}{8\,\lambda '}+\frac{45-240\log 2}{16\,\lambda '^{\,3/2}}  + O\left(\frac{1}{\lambda'^{\,2}}\right)\right) \notag \\[0.5em]
& + \frac{\lambda'}{4096}\rme^{-2\sqrt{\lambda'}} \left( 1+\frac{1+16\log 2}{\lambda '^{\,1/2}}-\frac{3-32\log 2-64(\log 2)^2}{\lambda '} + O\left(\frac{1}{\lambda'^{\,3/2}}\right)\right) \notag \\[0.5em]
& + O\left(\rme^{-3\,\sqrt{\lambda'}},\frac{1}{N^{2}} \right)
\label{strongF}
\end{align}
where the constant is given by
\begin{align}
k\,=\, - 3\log \mathsf{A} + \frac{1}{4} - \frac{11}{12}\log 2 + \frac{3}{4}\log(4\pi)
\end{align}
with $\mathsf{A}$ being the Glaisher constant and 
\begin{align}
\lambda' = \lambda\left(1-\frac{4\log 2}{\sqrt{\lambda}}\right)^2\equiv \lambda\,a(\lambda)^2\,.
\label{defa}
\end{align}
As noticed in \cite{Beccaria:2022ypy}, this coupling redefinition is advantageous as it sums up all the terms proportional to $\log 2$. The subleading terms in $1/\lambda'$ can be computed to any desired order. The expansion above provides direct access to the strong-coupling expansion of the Bremsstrahlung function. The final result will be presented explicitly in Section \ref{subsecfin}.

\subsubsection*{\textbf{D}-theory}

For this model the exact result for the Bremsstrahlung function is considerably more involved. From \eqref{BNLOD}-\eqref{BNNLOD}, it is evident that, in order to fully understand its strong-coupling behavior 
we need to analyze the following quantities:
\begin{align}
\mathsf{Y} \,, \qquad \int_0^\infty \!\mathcal{D}t\,\mathcal{Y}\left( \lambda,t \right)\,, \qquad \int_0^\infty\!\mathcal{D}t\,t\,\partial_t\mathcal{Y}\left( \lambda,t\right)\,, \label{quantities}
\end{align}
using the definitions in \eqref{Yhat}, \eqref{kernel} and \eqref{Ycal}. An efficient way to uniformly analyze the properties of these expressions at strong coupling is to introduce the integral function
\begin{align}
\mathcal{T}_{k,n}^{(p)}\left( \lambda \right)\equiv \int_0^\infty\!\frac{dt}{(t^2+\pi^2)^k}  \,
\frac{\rme^t\,t^{p}}{(\rme^t+1)^2}\,J_{n}\left(\frac{\sqrt{\lambda}\,t}{\pi}\right)\,,
\label{defI}
\end{align}
with $k$, $n$ and $p\,\in\mathbb{N}$. Indeed, the quantities \eqref{quantities} can be rewritten in terms of \eqref{defI} as follows
\begin{align}
\mathsf{Y}\,=\,& \frac{\sqrt{\lambda}}{\pi}\mathcal{T}_{0,1}^{(0)}(\lambda)-\frac{\log 2}{2\pi^2}\,\lambda\,, \label{Iquant} \\
\int_0^\infty \!\mathcal{D}t\,\mathcal{Y}\left( \lambda,t \right)\,=\,& \pi \sqrt{\lambda} \left( I_2(\sqrt{\lambda})\, \mathcal{T}_{1,1}^{(0)}(\lambda)+ \frac{1}{\pi} I_1(\sqrt{\lambda})\,\mathcal{T}_{1,2}^{(1)}(\lambda)  \right)\,, \label{IIquant} \\
\int_0^\infty\!\mathcal{D}t\,t\,\partial_t\mathcal{Y}\left( \lambda,t\right)\,=\,& \sqrt{\lambda}\,I_1(\sqrt{\lambda})\left(2\,\mathcal{T}_{1,2}^{(1)}(\lambda) -2\,\mathcal{T}_{2,2}^{(3)}(\lambda)+\frac{\sqrt{\lambda}}{2\pi}\,\mathcal{T}_{1,1}^{(2)}(\lambda) - \frac{\sqrt{\lambda}}{2\pi}\,\mathcal{T}_{1,3}^{(2)}(\lambda) \right) \notag \\ 
+ & \sqrt{\lambda}\,\pi\,I_2(\sqrt{\lambda})\left( \mathcal{T}_{1,1}^{(0)}(\lambda) -2\,\mathcal{T}_{2,1}^{(2)}(\lambda)+\frac{\sqrt{\lambda}}{2\pi}\,\mathcal{T}_{1,0}^{(1)}(\lambda) - \frac{\sqrt{\lambda}}{2\pi}\,\mathcal{T}_{1,2
}^{(1)}(\lambda)  \right) \,. \label{IIIquant} 
\end{align}
Therefore, in the following subsection we will study the strong-coupling behavior of the function $\mathcal{T}_{k,n}^{(p)}(\lambda)$, including both its perturbative and non-perturbative contributions. This analysis will allow us to derive the strong-coupling behavior of the quantities in \eqref{quantities}, and consequently of the Bremsstrahlung function in the \textbf{D}-theory.

\subsection{Perturbative and non-perturbative contributions}
\label{subsec:nonpert}

Our strategy consists in expressing the quantities $\mathcal{T}_{k,n}^{(p)}$ as an exact sum of an inverse power series in $\lambda$, which accounts for the perturbative sector at strong coupling, and an infinite series of modified Bessel functions of the second kind, $K_{\nu}(x)$, whose large-argument behaviour fully captures the non-perturbative corrections. Starting from the definition \eqref{defI}, we employ the partial fraction expansion of the $\text{sech}(x)$ function,
\begin{align}
\frac{\rme^t}{(\rme^t+1)^2}\,=\,\frac{1}{4}\text{sech}^2\left( \frac{t}{2} \right)\,=\,2\sum_{j=0}^\infty \frac{(2j+1)^2\pi^2-t^2}{\left(t^2+(2j+1)^2\pi^2 \right)^2} \,,
\label{sech}
\end{align}
obtaining 
\begin{align}  
\mathcal{T}_{k,n}^{(p)}\left( \lambda \right)= 2 \int_0^{\infty}\!dt\,t^{p}\,J_{n}\left(\frac{\sqrt{\lambda}\,t}{\pi}\right) \bigg[\,\frac{\pi^2-t^2}{\left(t^2+\pi^2 \right)^{k+2}} + \sum_{\substack{q=3 \\ q\;\text{odd}}}^\infty \frac{q^2\pi^2-t^2}{(t^2+\pi^2)^k\left(t^2+q^2\pi^2 \right)^2}\bigg]\,,
\label{defT1nn+1}
\end{align}
where $q= 2j+1$ and for convenience we have separated the $q=1$ case explicitly. Now we decompose the rational functions inside the square brackets into partial fractions as
\begin{align}
&\frac{\pi^2-t^2}{\left(t^2+\pi^2 \right)^{k+2}} = \frac{2\pi^2}{(t^2+\pi^2)^{k+2} }- \frac{1}{(t^2+\pi^2)^{k+1}} \,, \notag \\
& \frac{q^2\pi^2-t^2}{(t^2+\pi^2)^k\left(t^2+q^2\pi^2 \right)^2} = \frac{(-1)^k   \left((2 k-1) q^2+1\right)}{\pi ^{2 k} \left(q^2-1\right)^{k+1}(t^2+\pi ^2 q^2)}+ \frac{(-1)^k \, 2\,q^2 }{\pi ^{2 k-2}\left(q^2-1\right)^{k}\left(t^2+\pi ^2 q^2\right)^2}  \nonumber \\
& \qquad\qquad\qquad\qquad\qquad\qquad +\sum_{\ell=0}^{k-1} \frac{(-1)^{\ell}\left((2\ell+1)q^2+1\right)}{\pi^{2\ell+2}(q^2-1)^{\ell+2}(t^2+\pi^2)^{k-\ell}}\,.
  \label{simplefractions}
\end{align}
Plugging \eqref{simplefractions} into \eqref{defT1nn+1} we are left with the evaluation of integrals of the form 
\begin{align}
\int_0^{\infty}\frac{dt\;t^p}{(t^2+a^2)^m}\;J_n \left(\frac{t\sqrt{\lambda}}{\pi} \right) .
\label{genericint}
\end{align}
To simplify the discussion, we focus on the two families of functions relevant for our computation, namely $\mathcal{T}_{k,n}^{(n+1)}$ and $\mathcal{T}_{k,n}^{(n-1)}$, as one can check from \eqref{Iquant}-\eqref{IIIquant}. Since these two cases display qualitatively different behavior, we discuss them separately. More precisely, the former can be expressed exactly as a series of Bessel functions $K_{\nu}$ and receives no perturbative contributions at strong coupling, whereas the latter also includes a finite set of terms in the inverse power expansion of the coupling.

\subsubsection*{\texorpdfstring{The case $\mathcal{T}_{k,n}^{(n+1)}$}{}}
If we set $p=n+1$ in \eqref{genericint} we are able to perform the integrals of the fractions reported in \eqref{simplefractions} in full generality exploiting the identity  $(6.565,4)$ of \cite{gradshteyn2007}
\begin{align}
\int_0^{\infty}\!\frac{dt\,t^{n+1}}{(t^2+a^2)^{m}}\,J_{n}\!\left(\frac{\sqrt{\lambda}\,t}{\pi}\right)\,=\,\left(\frac{\sqrt{\lambda}}{2\pi}\right)^{m-1}\frac{a^{n-m+1}}{\Gamma(m)}K_{n-m+1}\!\left(\frac{\sqrt{\lambda}\,a}{\pi}\right)\,,
\label{Besselid}
\end{align}
valid for $\text{Re}(a)>0$ and $n<2m-\frac{1}{2}$, that yields to 
\begin{align}
\mathcal{T}_{k,n}^{(n+1)}\left( \lambda \right) &= \frac{ \lambda ^{\frac{k}{2}}}{2^{k-1}\,\pi ^{2k-n}\,\Gamma(k+2)} \left(\sqrt{\lambda }\, K_{k-n+1}(\sqrt{\lambda })-(k+1)\, K_{k-n}(\sqrt{\lambda })\right) \nonumber \\[4mm]
&+\frac{2\,(-1)^k}{\pi^{2k-n}}\sum_{\substack{q=3 \\ q\  \text{odd}}}^\infty     \bigg[\frac{ q^n}{\left(q^2-1\right)^{k+1}} \left(\left((2 k-1) q^2+1\right) K_n(q \sqrt{\lambda })+\sqrt{\lambda }\, q \left(q^2-1\right) K_{n-1}(q \sqrt{\lambda })\right) \nonumber\\
& +\sum_{\ell=0}^{k-1} (-1)^{\ell}
\frac{1+(2\ell+1)q^2}
{2^{\,k-\ell-2}\,\pi^{\,2k-n}\,(q^2-1)^{\ell+2}\,\Gamma(k-\ell)}
\,\lambda^{\frac{k-\ell-1}{2}}K_{k-\ell-n-1}(\sqrt{\lambda}) \bigg]\,.
\label{Tkn+1}
\end{align}
We remark that this result is exact in $\lambda$. For our purposes, we only need to evaluate it for four specific values of the parameters $n$ and $k$. The resulting expressions are explicitly reported in Appendix \ref{app:nonpert}. It is worth noting that, since in \eqref{Tkn+1} each term contains a Bessel function $K_\nu$, whose asymptotic expansion is given by
\begin{align}
    K_\nu(z) \underset{z \to \infty}{\sim}  \left(\frac{\pi}{2 z}\right)^{\frac{1}{2}}\,e^{-z} \sum_{j=0}^{\infty} \frac{a_j(\nu)}{z^j} 
\label{asymK}
\end{align}
with $a_j(\nu)$ defined in \eqref{aj}, the quantity $\mathcal{T}_{k,n}^{(n+1)}$ receives, in the limit $\lambda \to \infty$, only non-perturbative, exponentially suppressed corrections. This is also confirmed by the Mellin-Barnes approach shown in Appendix \ref{app:pert}.

\subsubsection*{\texorpdfstring{The case $\mathcal{T}_{k,n}^{(n-1)}$}{}}
Setting $p=n-1$ in \eqref{genericint}, we can evaluate the integrals building on the formula (6.565,8) of \cite{gradshteyn2007} and explicitly deriving the following identity
\begin{align}
\int_0^\infty \frac{dt\,t^{n-1} }{\left(t^2 + a^2\right)^m} J_{n}\bigg(\frac{\sqrt{\lambda}\,t}{\pi}\bigg)  =& \frac{2^{n-1} \pi ^{n} \Gamma (n)}{a^{2m} \lambda ^{\frac{n}{2} }} +\frac{(-1)^{m}}{(m-1)!}\left[ \partial^{m-1}_s\,h_n(s)\right]_{s=a^2}, \label{id1} 
\end{align}
where 
\begin{align}
  \left[ \partial^{m-1}_s\,h_n(s)\right]_{s=a^2} \equiv & \left[ \partial^{m-1}_s\left(s^{\frac{n}{2}-1} K_{n}\!\left(\frac{\sqrt{\lambda\,s}}{\pi }\right)\right)\right]_{s=a^2} \notag \\[0.5em]
  = &(-1)^{m-1}a^{n-2m}
 \sum_{i=0}^{m-1}\frac{(m-1)!}{i!\,2^i} \left( \frac{\sqrt{\lambda}\,a}{\pi}\right)^i K_{n-i}\! \left( \frac{\sqrt{\lambda}\,a}{\pi} \right)\,,
\end{align}
which is valid for $\text{Re}(a)>0$ and $n<2m+\frac{3}{2}$. 
Therefore, after inserting \eqref{simplefractions} in \eqref{defT1nn+1} and then using \eqref{id1}, we finally find
\begin{comment}

\end{comment}
\begin{align}
&\mathcal{T}_{k,n}^{(n-1)} = \frac{2^n\,\Gamma(n)}{\pi^{2k-n+2}\,\lambda^{\frac{n}{2}}}\left( 1+\sum_{\substack{q=3 \\ q\  \text{odd}}}^\infty \frac{(-1)^k\left((2 k+1) q^2-1\right)}{q^2(q^2-1)^{k+1}} +\!\sum_{\substack{q=3 \\ q\  \text{odd}}}^\infty\sum_{\ell=0}^{k-1} \frac{(-1)^{\ell}\left((2 \ell+1) q^2+1\right)}{\left(q^2-1\right)^{\ell+2}} \right) \nonumber \\
&  + \frac{2\,(-1)^k}{k!}\! \bigg[ \!\!\left(\frac{2\pi^2}{k+1}\partial_s^{k+1}+ \partial_s^k\right)\! h_n(s) \bigg]_{s=\pi^2} \!\!+\!\!\sum_{\substack{q=3 \\ q\  \text{odd}}}^\infty\!\sum_{\ell=0}^{k-1} \frac{ 2\,(-1)^{k}\left((2 \ell+1) q^2+1\right)}{ \pi ^{2\ell+2} \left(q^2-1\right)^{\ell+2}(k-\ell-1)!} \!\left[\partial_s^{k-\ell-1}\,h_n(s) \right]_{s=\pi^2} \nonumber \\
&-\sum_{\substack{q=3 \\ q\  \text{odd}}}^\infty \frac{2\,(-1)^k\,q^{n-2}}{\pi^{2k-n+2}(q^2-1)^{k+1}}\bigg( \left((2 k+1) q^2-1\right) K_n(q\sqrt{ \lambda})+\left(q^3-q\right)\sqrt{\lambda}\,K_{n-1}(q\sqrt{ \lambda })\bigg) \,.
\label{Tkn-1}
\end{align}
We observe that the term appearing in the first line coincides exactly with the perturbative strong-coupling contribution, which consists of a single term in the inverse power expansion in $\lambda$, as also confirmed by the alternative method in Appendix \ref{app:pert}. The remaining terms in \eqref{Tkn-1}, involving modified Bessel functions, encode the non-perturbative corrections. In Appendix \ref{app:nonpert}, we present the explicit form of \eqref{Tkn-1} for the four integrals relevant to our computation.

\subsection{Final results}
\label{subsecfin}

Exploiting the results derived in the previous subsection, we can now present the final strong-coupling expansions for the Bremsstrahlung function in the \textbf{E}- and \textbf{D}-theories. For convenience, we introduce
\begin{align}
\Delta B_\mathbf{E}\,\equiv\,& B_\mathbf{E}-B_{\mathcal{N}=4}\,, \label{dE} \\
\Delta B_\mathbf{D}\,\equiv\,& B_\mathbf{D} - B_\mathbf{E}\,. \label{dD}
\end{align}
From \eqref{BLOE} it immediately follows
\begin{align}
\Delta B_\mathbf{E}^{(0)}\,=\,\Delta B_\mathbf{E}^{(1)}\,=\,0\,,
\end{align}
while, by plugging \eqref{strongF} in \eqref{BNLOE}, for the next contribution in the large-$N$ limit it is straightforward to get
\begin{align}
\Delta B_\mathbf{E}^{(2)}  \underset{\lambda \to \infty}{\sim} & \,\frac{1}{8\,\pi^2} \biggl[-\frac{\lambda^{3/2}}{16} + \frac{3\,\lambda}{8}-\frac{3\,\log 2\,\lambda^{1/2}}{8\,a(\lambda)}-\frac{45\,\zeta(3)}{128\,\lambda^{1/2}\,a(\lambda)^4} -\frac{4725\,\zeta(5)}{1024\,\lambda^{3/2}\,a(\lambda)^6} +O\!\left(\frac{1}{\lambda^{2}\,a(\lambda)^7}\right)  \notag \\[0.5em]
&  \!\!\! + \frac{\ii\,\lambda^2\,a(\lambda)}{128}\rme^{-a(\lambda)\sqrt{\lambda}} \left( 1-\frac{1-16\log 2}{2\,\lambda^{1/2}\,a(\lambda)}-\frac{15-96\log 2}{8\,\lambda\,a(\lambda)^2}+\frac{15-48\log 2}{16\,\lambda^{3/2}\,a(\lambda)^3} +O\!\left(\frac{1}{\lambda^{2}\,a(\lambda)^4}\right) \right) \notag \\[0.5em]
& \!\!\! + \frac{\lambda^{5/2}\,a(\lambda)^2}{4096}\rme^{-2a(\lambda)\sqrt{\lambda}} \left( 1+\frac{16\log 2}{\lambda^{1/2}\,a(\lambda)}-\frac{7-48\log 2-128(\log 2)^2}{2\,\lambda\,a(\lambda)^2} +O\left(\frac{1}{\lambda^{3/2}\,a(\lambda)^3}\right) \right)\biggr] \notag \\[0.5em]
& \!\!\! + O\!\left(\rme^{-3\,a(\lambda)\sqrt{\lambda}} \right),
\label{strongdE}
\end{align}
where $a(\lambda)$ is defined in \eqref{defa}. As explained in \cite{Bajnok:2024epf,Bajnok:2024ymr}, the imaginary coefficient of the first non-perturbative correction is exactly cancelled by the imaginary terms generated when circumventing the Borel singularities of the perturbative expansion in the first line of \eqref{strongdE}. The above expression concludes our analysis of the Bremsstrahlung function in the \textbf{E}-theory.

In the \textbf{D}-theory the planar equivalence for the Bremsstrahlung function continues to hold, as shown in \eqref{BLOD}, implying that
\begin{align}
\Delta B_\mathbf{D}^{(0)}\,=\,0\,.
\end{align}
Nevertheless, starting from the next two orders in the large-$N$ expansion, the structure becomes significantly more involved, as proven in \eqref{BNLOD} and \eqref{BNNLOD}. Relying on the detailed analysis presented in the previous subsection, however, we are able to derive the explicit strong-coupling expansions for these subleading large-$N$ contributions, including both perturbative and non-perturbative corrections. In particular, substituting \eqref{1t}-\eqref{8t} into \eqref{Iquant}-\eqref{IIIquant} allows us to express \eqref{BNLOD} and \eqref{BNNLOD} entirely in terms of modified Bessel functions, \(I_\mu\) and \(K_\mu\). We then use the large-argument asymptotic expansions of these functions, given in \eqref{expBI} and \eqref{asymK}, to obtain
\begin{align}
    \Delta B_{\mathbf{D}}^{(1)} \underset{\lambda \to \infty}{\sim}& \,\frac{1}{16\,\pi^4}\left(-\log 2\,\lambda ^{3/2}+\frac{(4\, \pi ^2+3\,\log 2)\lambda^{1/2}}{8}+ \frac{3\, \log 2}{4}-\frac{3 (8 \pi ^2-63\, \log 2)}{128  \, \lambda^{1/2} }+O\!\left(\frac{1}{\lambda}\right) \right) \nonumber \\[0.5em] 
    & \!\!\! + \frac{\lambda ^{9/4}\,\text{e}^{-\sqrt{\lambda}}}{24\, \sqrt{2}\, \pi^{7/2}}\left(1-\frac{33}{8\,\lambda^{1/2}}+\frac{64 \,\pi ^2-519}{128 \,\lambda } -\frac{9 (64\, \pi ^2+515)}{1024\, \lambda ^{3/2}}+  \frac{9 (128 \, \pi ^2-3037)}{32768 \, \lambda ^2} +O\!\left(\frac{1}{\lambda^{5/2}}\right) \right)\nonumber\\[0.5em]
    & \!\!\! + \frac{\text{i}\,\lambda^2\, \text{e}^{-2\sqrt{\lambda}}}{4 \,\pi^4}\left( \log 2+\frac{\log 2}{4\,\lambda^{1/2}}-  \frac{16 \,\pi^2-21 \log 2}{32 \,\lambda }-\frac{16 \,\pi^2-129 \log 2}{128\, \lambda^{3/2}}+O\!\left(\frac{1}{\lambda^{2}}\right)\right) \nonumber \\[0.5em]
     & \!\!\! + O\!\left(\text{e}^{-3\sqrt{\lambda}}\right)\,, \label{finBD1} \\[1em]
    \Delta B_{\mathbf{D}}^{(2)} \underset{\lambda \to \infty}{\sim}& \,\frac{3  (\log 2)^2\lambda ^{5/2}}{128\, \pi ^6}-\frac{ \log 2 \left(8 \,\pi ^2+3\log 2\right)\!\lambda ^{3/2}}{1024 \,\pi ^6}-\frac{ \left(32 \pi ^4+48\, \pi ^2 \log 2 -189 (\log 2)^2\right)\!\lambda^{1/2}}{16384 \,\pi ^6}  +O\!\left(\lambda^0\right) \nonumber \\[0.5em] 
    &\!\!\!+\frac{\lambda ^{15/4} \,\text{e}^{-\sqrt{\lambda}}}{96\, \sqrt{2}\, \pi ^{11/2}}  \left(\log 2-\frac{69 \log 2}{8\, \lambda^{1/2} } -\frac{64 \pi ^2 -\log 2(64\,\pi^2+1329) }{128\, \lambda }+O\!\left(\frac{1}{\lambda^{3/2}}\right) \right)\nonumber \\[0.5em]
    &\!\!\!+\frac{\lambda ^4\,\text{e}^{-2\sqrt{\lambda}}}{96 \,\pi ^5} \left(1 -\frac{ 11 \,\pi-24\, \text{i} (\log 2)^2}{4 \,\pi  \,\lambda^{1/2} } +\frac{64 \,\pi ^3-729 \,\pi -1008 \,\text{i} (\log 2)^2}{96 \,\pi \, \lambda }+O\!\left(\frac{1}{\lambda^{3/2}}\right) \right)\nonumber  \\[0.5em]
    &\!\!\! +O\!\left(\text{e}^{-3\sqrt{\lambda}}\right)\,. \label{finBD2}
\end{align}
The contributions corresponding to the more strongly exponentially suppressed factors can be immediately derived from the exact expressions given in terms of infinite sums of Bessel functions $K_\mu$. Moreover, the series in inverse powers of the coupling multiplying each exponential can be systematically extended to arbitrary order. In both \eqref{finBD1} and \eqref{finBD2}, the imaginary terms arising in the $\rme^{-2\sqrt{\lambda}}$ sector stem from the transseries parameter $\ii\rme^{\mu\,\pi\,\ii}$ entering the asymptotic expansion of the modified Bessel function of the first kind. On the other hand, the terms involving the functions $\mathcal{T}_{k,n}^{(p)}$ give rise to no imaginary contributions, as they are built from sums of modified Bessel functions of the second kind. 

The $\log 2$-terms appearing in the strong-coupling expansions \eqref{finBD1} and \eqref{finBD2} may be absorbed into a redefinition of the ’t Hooft coupling by introducing the shifted variable $\widehat{\lambda}$ \cite{Beccaria:2022kxy,Billo:2024ftq,Billo:2026lzb}
\begin{align}
\frac{1}{\widehat{\lambda}}=\frac{1}{\lambda}+\frac{\log 2}{2\pi^2 N}\,.
\label{lambdahat}
\end{align}
Notice that this coupling redefinition is qualitatively different from the one used for the \textbf{E}-theory in \eqref{defa}. Since $\widehat{\lambda}$ depends explicitly on $N$, it mixes different orders in the $1/N$ expansion, unlike the $N$-independent redefinition appearing in the strong-coupling expansions \eqref{strongdE}\,\footnote{This difference has a direct matrix model origin. In the
\textbf{D}-theory, the $\log 2$-terms
originate from the single-trace interaction action generated by the fundamental
hypermultiplets. This interaction contributes already at order $1/N$, while
its repeated insertions enter at successive orders in the large-$N$
expansion, naturally leading to an $N$-dependent effective coupling. In the
\textbf{E}-theory, by contrast, the $\log 2$-dependence emerges from the
strong-coupling expansion of the free energy at a fixed order
in large-$N$ limit, and therefore leads to an $N$-independent redefinition.}. For the cancellation of all $\log 2$-terms to be exact, one must first express $B^{(0)}_{\mathbf D}$ in terms of $\widehat{\lambda}$, and subsequently recast the expansions \eqref{finBD1} and \eqref{finBD2} in the same variable. With this choice of effective coupling, the final expressions become significantly more compact:
\begin{align}
    \Delta B_{\mathbf{D}}^{(1)} \underset{\widehat{\lambda}\to \infty}{\sim}&\, \frac{1}{32\,\pi^2}\left(\widehat{\lambda}^{\,1/2}-\frac{3}{8\,\widehat{\lambda}^{\,1/2}}-\frac{3}{4\,\widehat{\lambda}}-\frac{189}{128\,\widehat{\lambda}^{\,3/2}}-\frac{27}{8\,\widehat{\lambda}^{2}}+O\!\left(\frac{1}{\widehat{\lambda}^{\,5/2}}\right)\! \right) \nonumber \\[0.5em] 
    &  \!\!\!\!\!\!\!\!\!  + \frac{\widehat{\lambda} ^{9/4}\,\text{e}^{-\sqrt{\widehat{\lambda}}}}{24\, \sqrt{2}\, \pi^{7/2}}\left(1-\frac{33}{8\,\widehat{\lambda}^{\,1/2}}+\frac{64 \,\pi ^2-519}{128 \,\widehat{\lambda} } -\frac{9 (64\, \pi ^2+515)}{1024\, \widehat{\lambda} ^{\,3/2}}+  \frac{9 (128 \, \pi ^2-3037)}{32768 \, \widehat{\lambda} ^2} +O\!\left(\frac{1}{\widehat{\lambda}^{\,5/2}}\right) \!\right) \notag \\[0.5em]
    & \!\!\!\!\!\!\!\!\! -\frac{\text{i}\, \widehat{\lambda}\,\text{e}^{-2\sqrt{\widehat{\lambda}}}}{8 \,\pi ^2}\left( 1+\frac{1}{4 \widehat{\lambda}^{1/2}} +\frac{21}{32 \widehat{\lambda}}+\frac{129}{128 \widehat{\lambda}^{3/2}} +\frac{3147}{2048 \widehat{\lambda}^2}+O\!\left(\frac{1}{\widehat{\lambda}^{\,5/2}}\right) \!\right) + O\!\left(\text{e}^{-3\sqrt{\widehat\lambda}}\right) \,, \label{fin2BD1} \\[1em]
    \Delta B_{\mathbf{D}}^{(2)} \underset{\widehat{\lambda}\to \infty}{\sim} & \, \frac{1}{512\,\pi^2}\left( -\widehat{\lambda}^
    {\,1/2}+\frac{9}{8\, \widehat{\lambda}^{\,1/2}}+ \frac{3}{\widehat{\lambda}}+\frac{945}{128\, \widehat{\lambda}^{\,3/2}}+\frac{81}{4\, \widehat{\lambda }^2}+O\!\left(\frac{1}{\widehat{\lambda}^{\,5/2}}\right) \!\right) \nonumber \\[0.5em] 
    &\!\!\!\!\!\!\!\!\! -\frac{\widehat{\lambda}^{11/4}\,\text{e}^{- \sqrt{\widehat{\lambda}}}}{192\, \sqrt{2} \,\pi ^{7/2}} \!\left(\!1-\frac{53}{8 \widehat{\lambda}^{1/2}} +\frac{64 \pi ^2+273}{128 \widehat{\lambda}}- \frac{832 \pi ^2+2559}{1024\, \widehat{\lambda}^{3/2}} -\frac{63 \left(128 \pi ^2+1611\right)}{32768 \,\widehat{\lambda}^2} +O\!\left(\frac{1}{\widehat{\lambda}^{\,5/2}}\right) \!\right) \nonumber \\[0.5em] 
    &\!\!\!\!\!\!\!\!\! +\frac{\widehat{\lambda}^4\,\text{e}^{-2\sqrt{\widehat{\lambda}}}}{96\, \pi ^5}\!\left(\!1-\frac{11}{4 \,\widehat{\lambda}^{1/2}}+\frac{64 \pi ^2-729}{96\, \widehat{\lambda}}-\frac{64 \pi ^2+603}{128\, \widehat{\lambda}^{3/2}}+\frac{512 \pi ^4-3456 \pi ^2-14607}{6144\, \widehat{\lambda}^2} +O\!\left(\frac{1}{\widehat{\lambda}^{\,5/2}}\right) \!\right) \nonumber \\[0.5em] 
    &\!\!\!\!\!\!\!\!\! + O\!\left(\text{e}^{-3\sqrt{\widehat\lambda}}\right) \,. \label{fin2BD2}
\end{align}
It is useful to organize these large-$\widehat{\lambda}$ expansions into their perturbative and exponentially suppressed sectors as follows\,\footnote{We emphasize once again that, by instantons, here we mean contributions exponentially suppressed in $\widehat{\lambda}$.} 
\begin{align}
    \Delta B_{\mathbf{D}}^{(n)} = \Delta B_{\mathbf{D}}^{(n)}\bigg|_{\text{pert}} + \text{e}^{-\sqrt{\widehat{\lambda}}}\, \Delta B_{\mathbf{D}}^{(n)}\bigg|_{1\text{-inst}} + \text{e}^{-2\,\sqrt{\widehat{\lambda}}}\, \Delta B_{\mathbf{D}}^{(n)}\bigg|_{2\text{-inst}}\,.
\end{align}
This decomposition reveals a simple sector-wise relation between the $n=1$ and $n=2$ results.  In particular, the perturbative and
one-instanton sectors of $\Delta B_{\mathbf{D}}^{(2)}$ can be obtained from the corresponding sectors of $\Delta B_{\mathbf{D}}^{(1)}$ by the
following differential operators\,\footnote{The symbol $=$ in the two following equations is to be understood in the large-$\widehat{\lambda}$ limit.}:
\begin{align}
    \Delta B_{\mathbf{D}}^{(2)}\bigg|_{\text{pert}} &= \frac{1}{8} \left( \widehat{\lambda} \,\partial_{\widehat{\lambda}} - 1 \right) \Delta B_{\mathbf{D}}^{(1)}\bigg|_{\text{pert}} \,, \\[0.5em]
\text{e}^{-\sqrt{\widehat{\lambda}}}\,\Delta B_{\mathbf{D}}^{(2)}\bigg|_{1\text{-inst}} &= \frac{1}{4} \left( \widehat{\lambda} \,\partial_{\widehat{\lambda}} - 1 \right) \left[\text{e}^{-\sqrt{\widehat{\lambda}}}\,\Delta B_{\mathbf{D}}^{(1)}\bigg|_{1\text{-inst}}\right] \,.
\end{align}
The above relations do not extend in a straightforward way to the two-instanton sector. The mismatch already appears at the level of the
leading power in $\widehat\lambda$, indicating that the two-instanton contribution of $\Delta B_{\mathbf D}^{(2)}$ is not generated from that of $\Delta B_{\mathbf D}^{(1)}$ by the same simple differential operator. This behavior can be understood from the non-linear structure of the exact
result \eqref{BNNLOD}, which originates from the large-$N$ expansion of the
ratio $\langle W'\rangle/\langle W\rangle$. 
At order $\rme^{-2\sqrt{\widehat{\lambda}}}$, the expansion contains products of the $\rme^{-\sqrt{\widehat{\lambda}}}$ contributions
to $W_{\mathbf D}^{(1)}$ and $W_{\mathbf D}'{}^{(1)}$, as well as the square
of the corresponding contribution to $W_{\mathbf D}^{(1)}$. Such terms are not determined by the
second exponential sector of $\Delta B_{\mathbf D}^{(1)}$ alone. Therefore, the sector-wise linear
relation need not extend to this order. However, presumably, non-linear
identities may still point to an underlying structure for the strong-coupling expansion of the Bremsstrahlung function, relating different orders in the $1/N$ expansion to one another. This could provide a more efficient way to access the full tower of $1/N$ corrections, whose
direct determination from localization becomes increasingly involved. At present, however, the available strong-coupling expansions are not sufficient
to establish whether such a structure persists to all sub-planar orders.

\section{Conclusions}
\label{sec:concl}

In this paper, we have computed the Bremsstrahlung function up to the first three orders in the large-$N$ expansion for two distinct $\mathcal{N}=2$ superconformal gauge theories with gauge group $\mathrm{SU}(N)$, the \textbf{E}- and \textbf{D}-theories. By employing matrix model techniques derived from supersymmetric localization, we have obtained exact results, valid for arbitrary values of the coupling, at each large-$N$ order. We also provided a detailed analysis of the strong-coupling regime, deriving the complete series of corrections in inverse powers of the 't Hooft coupling, together with closed-form expressions also for the exponentially suppressed contributions. 

A natural extension of the results presented in this paper is the analysis of the Bremsstrahlung function in a different $\mathcal{N}=2$ theory with gauge group $\mathrm{Sp}(N)$, considered for instance in \cite{Beccaria:2021ism,Beccaria:2022kxy,DeLillo:2025hal}, which closely resembles the \textbf{D}-theory. However, this model admits an easier matrix model description, involving only a single-trace interaction action. This simplified structure has led to significant progress in matrix model computations: in particular, it has been shown that several observables satisfy a Toda lattice equation, enabling to derive recursive relations that control the full $1/N$ expansions \cite{Beccaria:2022kxy,DeLillo:2025hal}. In this theory, it should therefore be more accessible to compute many more subleading corrections beyond the planar limit. 

A second interesting direction would be to analyze the Bremsstrahlung function in the very strong-coupling limit, obtained by sending the number of colors $N$ to infinity while keeping the gauge coupling $g$ fixed, rather than the 't Hooft coupling $\lambda$. In this regime, Yang-Mills instanton effects cannot be neglected, and their contributions are essential for shedding light on the modular properties of these theories. Indeed, the implications of S-duality are highly non-trivial for extended objects, such as Wilson loops and, consequently, the cusped Wilson loops considered in this paper. An important recent study in this direction was carried out in \cite{Dorigoni:2024vrb}, where the large-$N$ instanton contributions to correlation functions of local operators in the presence of a half-BPS line defect have been thoroughly investigated in $\mathcal{N}=4$ SYM theory. Nonetheless, extending these calculations also to the $N\rightarrow\infty$ limit of the $\mathcal{N}=2$ setups studied here is far from straightforward, due to the intricate structure of the corresponding matrix models. This remains an open problem that would be important to address.

Finally, our results for the Bremsstrahlung function in the strong-coupling regime yield precise predictions for the holographic duals of the models under consideration. The radiation emitted by an accelerated heavy external quark in $\mathcal{N}=4$ SYM was first computed holographically in \cite{Mikhailov:2003er}. The quark, transforming in the fundamental representation of the gauge group, is described in the dual geometry by a fundamental open string in AdS$_5$, with an endpoint attached to the AdS boundary. Mikhailov constructed the nonlinear wave solution induced on the string worldsheet by the boundary motion and computed the associated energy flux. In this sense, the Bremsstrahlung function admits a direct worldsheet interpretation as the coefficient controlling the radiative energy flux carried by outgoing nonlinear waves on the fundamental string, whose boundary endpoint follows the accelerated trajectory of the external quark. The $\mathcal{N}=2$ gauge theories studied in this work have holographic duals obtained through orbifold/orientifold projections acting on the internal space of $\mathrm{AdS}_5\times S^5$. We have shown that, in both models, the Bremsstrahlung function exhibits planar equivalence with its $\mathcal{N}=4$ SYM counterpart. This suggests that, in the corresponding holographic setups, one should recover the same result derived in \cite{Mikhailov:2003er}. However, that computation is valid only in the supergravity regime, obtained by taking both $N\to\infty$ and $\lambda\to\infty$.
Using matrix-model techniques, together with the detailed strong-coupling analysis carried out in Section~\ref{sec:strong}, we were able to compute both higher-genus corrections and the complete series of higher-derivative contributions to the bulk effective action. Moreover, we carefully evaluated the exponentially suppressed terms appearing in the strong-coupling regime, which, according to the holographic dictionary, correspond to effects of the form
\begin{align}
\rme^{-n\sqrt{\lambda}}
\,=\,
\rme^{-\frac{n L^2}{\alpha'}}
\,,\qquad n\in\mathbb{N}^+ \, ,
\end{align}
where $L$ denotes the radius of $\mathrm{AdS}_5$. On the gravity side, these exponential terms can be interpreted as instanton weights associated with worldsheet instanton contributions, while the perturbative expansion around each exponential captures fluctuations about the corresponding classical worldsheet solution (see, for instance, \cite{Drukker:2006ga}).
Despite the presence of the additional D$7$-branes in the holographic dual of the $\mathbf{D}$-theory, the non-perturbative corrections are governed by the same leading exponential scale $\rme^{-n\sqrt{\lambda}}$ as in the $\mathbf{E}$-theory. In the worldsheet instanton perspective, this indicates that the corresponding instanton actions have the same leading scaling with the string tension. The series in inverse powers of $\sqrt{\lambda}$ multiplying these exponentials is nevertheless different, even at fixed order in $1/N$. This may reflect the additional open string sector, since the D$7$-branes may modify the fluctuation determinants of the relevant instanton configurations. A direct string theory analysis would be required to test this possibility and clarify the origin of the different fluctuation series.

\vskip 1cm
\noindent {\large {\bf Acknowledgments}}
\vskip 0.2cm
We would like to thank Marco Billò, Marialuisa Frau, Francesco Galvagno, Alberto Lerda, Alessandro Pini, Christoph Uhlemann and Congkao Wen for useful discussions and for helpful comments on the draft. 

This research is partially supported by the INFN project ST\&FI ``String Theory \& Fundamental Interactions''. 

\vskip 1cm

\appendix

\section{Explicit expressions for non-perturbative contributions}
\label{app:nonpert}
In this appendix, we provide the explicit form of the integrals needed in our computation, expressed in terms of modified Bessel functions of the second kind, $K_\mu(x)$.
Exploiting \eqref{Tkn+1} we get\,\footnote{For $\mathcal{T}_{2,2}^{(3)}$ we need to use the equation \eqref{Besselid} also for $m=1$ and $n=2$. In this case it should be understood via analytic continuation.} 
\begin{align}
    &\mathcal{T}_{1,0}^{(1)}\left( \lambda \right) =\frac{6\lambda+\pi^2-3}{12\pi^2}\,K_0(\sqrt{\lambda }) -\frac{2}{\pi^2}\sum_{\substack{k=3 \\ k\  \text{odd}}}^\infty \left[\frac{k\,\sqrt{\lambda}}{k^2-1}\,K_1(k\sqrt{\lambda})+\frac{k^2+1}{(k^2-1)^2}\,K_0(k\sqrt{\lambda})  \right] \,, \label{1t} \\[2mm]
     &\mathcal{T}_{1,1}^{(2)}=\frac{6 \lambda +\pi ^2-3 }{12 \pi}K_1(\sqrt{\lambda })-\frac{ \sqrt{\lambda } }{\pi}K_0(\sqrt{\lambda }) \nonumber \\[2mm] 
    & \quad \quad \quad \quad  -\frac{2}{\pi}\sum_{\substack{k=3 \\ k\  \text{odd}}}^\infty   \frac{k}{\left(k^2-1\right)^2}   \bigg[\left(k^2+1\right) K_1(k \sqrt{\lambda })+k \left(k^2-1\right) \sqrt{\lambda } K_{0}(k \sqrt{\lambda }) \bigg] \,, \\[2mm]
    &\mathcal{T}_{2,1}^{(2)}=\frac{1}{24\pi^3}\left[\sqrt{\lambda } \left(2 \lambda +\pi ^2-3\right) K_0(\sqrt{\lambda })-(2 \lambda +3) K_1(\sqrt{\lambda })\right]& \nonumber \\[2mm]
    &\quad \quad \quad \quad +\frac{2}{\pi^3} \sum_{\substack{k=3 \\ k\  \text{odd}}}^\infty \frac{ k}{(k^2-1)^3} \left[\left(3 k^2+1\right) K_1(k \sqrt{\lambda })+k \left(k^2-1\right) \sqrt{\lambda } K_{0}(k \sqrt{\lambda })\right]\,,\\[2mm]
    &\mathcal{T}_{2,2}^{(3)}=\frac{1}{24\pi^2}\left[\sqrt{\lambda } \left(2 \lambda +\pi ^2+9\right) K_1(\sqrt{\lambda })-3 (2 \lambda +1) K_2(\sqrt{\lambda })\right]& \nonumber \\[2mm]
    & \quad \quad \quad \quad +\frac{2}{\pi^2} \sum_{\substack{k=3 \\ k\  \text{odd}}}^\infty \frac{ k^2}{(k^2-1)^3} \left[\left(3 k^2+1\right) K_2(k \sqrt{\lambda })+k \left(k^2-1\right) \sqrt{\lambda } K_{1}(k \sqrt{\lambda })\right]\,.
\end{align}
Instead, using \eqref{Tkn-1}, we obtain 
\begin{align}
&\mathcal{T}_{0,1}^{(0)}= \frac{\pi }{4 \sqrt{\lambda }} -\frac{2}{\pi} \sum_{\substack{k=1 \\ k\  \text{odd}}}^\infty \frac{1}{k} \left[ K_{1}(k \sqrt{\lambda })+k \sqrt{\lambda }\, K_0(k \sqrt{\lambda })\right]\,, \\[2mm]
&\mathcal{T}_{1,1}^{(0)}= \frac{1}{4 \pi  \sqrt{\lambda }}-\frac{6 \lambda +\pi ^2+21 }{12 \pi ^3}K_1(\sqrt{\lambda })- \frac{\sqrt{\lambda}}{\pi^3}K_0(\sqrt{\lambda}) \nonumber \\[2mm]
& \quad \quad \quad \quad+ \frac{2}{\pi^3}\sum_{\substack{k=3 \\ k\  \text{odd}}}^\infty \frac{ 1}{ k \left(k^2-1\right)^2}\bigg[\left(k^2+1\right) K_1(k \sqrt{\lambda })+k (k^2-1)\sqrt{\lambda }\, K_2(k \sqrt{\lambda }) \bigg]\,, \\[2mm]
&\mathcal{T}_{1,2}^{(1)}= \frac{1}{2 \lambda }-\frac{6 \lambda +\pi ^2+21}{12 \pi ^2}\ K_2(\sqrt{\lambda }) \nonumber \\[2mm]
&\quad \quad \quad \quad + \frac{2}{\pi^2}\sum_{\substack{k=3 \\ k\  \text{odd}}}^\infty \frac{1}{ \left(k^2-1\right)^2}\bigg[\left(-k^2+3\right) K_2(k \sqrt{\lambda })+k \left(k^2-1\right) \sqrt{\lambda }\, K_3(k \sqrt{\lambda }) \bigg]\,,\\[2mm ] 
&\mathcal{T}_{1,3}^{(2)}= \frac{2 \pi }{\lambda ^{3/2}}-\frac{6 \lambda +\pi ^2+21}{12 \pi} K_3(\sqrt{\lambda }) +\frac{ \sqrt{\lambda }}{\pi}K_2(\sqrt{\lambda })\nonumber \\[2mm]
& \quad \quad \quad \quad + \frac{2}{\pi}\sum_{\substack{k=3 \\ k\  \text{odd}}}^\infty \frac{k}{(k^2-1)^2} \bigg[ \left(-3k^2+5\right) K_3(k \sqrt{\lambda })+k \left(k^2-1\right) \sqrt{\lambda } \,K_4(k \sqrt{\lambda })\bigg]\label{8t} \,.
\end{align}

\section{Different method for the perturbative series at strong coupling}
\label{app:pert}

In this appendix, we present an alternative method for extracting the perturbative contributions at strong coupling, namely the terms in the expansion in inverse powers of the ’t Hooft coupling. This approach relies on the Mellin-Barnes integral representation of the Bessel functions, 
\begin{align}
J_n\left(\frac{\sqrt{\lambda}\,t}{\pi}\right)
\,=\,
\int_{-\ii\infty}^{\ii\infty}\frac{ds}{2\pi\ii}\,
\frac{\Gamma\left(-s \right)}{\Gamma\left(s+n+1\right)}
\left(\frac{\sqrt{\lambda}\,t}{2\pi}\right)^{2s+n}\,,
\end{align}
through which the functions \eqref{defI} can be recast as
\begin{align}
\mathcal{T}_{k,n}^{(p)}\left( \lambda \right)=\int_{-\ii\infty}^{\ii\infty}\frac{ds}{2\pi\ii}\,\frac{\Gamma\left(-s \right)}{\Gamma\left(s+n+1\right)}\left(\frac{\sqrt{\lambda}}{2\pi}\right)^{2s+n}\int_0^\infty\!\frac{dt}{(t^2+\pi^2)^k}  \,
\frac{\rme^t\,t^{2s+n+p}}{(\rme^t+1)^2}\,.
\end{align}
Following an analysis similar to that of \cite{Pufu:2023vwo}, the integral over \(t\) can be performed explicitly. For this purpose, we introduce the functions
\begin{align}
\mathcal{J}_k(s)
\,=\,
\int_0^\infty \frac{dt}{(t^2+\pi^2)^k} \,
\frac{\rme^t\,t^{2s+3}}{(\rme^t+1)^2}\,,
\label{defJ}
\end{align}
which are defined for \(\mathrm{Re}(s)>-2\). For $k=0$ we have
\begin{align}
\mathcal{J}_0(s)\,=\,
\int_0^\infty dt \,
\frac{\rme^t}{(\rme^t+1)^2}\,t^{2s+3}
\,=\,\Gamma(2s+4)\,\eta(2s+3)\,,
\label{J0}
\end{align}
where \(\eta(s)\) denotes the Dirichlet eta function. For generic $k$ they satisfy the recurrence relations
\begin{align}
&\mathcal{J}_1(s)+\pi^2\mathcal{J}_1(s-1)
\,=\,\mathcal{J}_0(s-1)\,=\,
\Gamma(2s+2)\,\eta(2s+1)\,, \label{recurrence1}\\
& \mathcal{J}_k(s)+ \pi^2 \mathcal{J}_k(s-1)= \mathcal{J}_{k-1}(s-1) \quad \text{for}\ k\geq 2\,. 
\label{recurrence2}
\end{align}
These relations, valid for \(\mathrm{Re}(s)>-1\), provide a meromorphic continuation of \(\mathcal{J}_k(s)\) to the whole complex plane. After continuation, \(\mathcal{J}_k(s)\) are regular at \(s=-1\) and have simple poles at \(s=-2,-3,\dots\). Using \eqref{defJ}, we can rewrite \eqref{defI} as
\begin{align}
\mathcal{T}_{k,n}^{(p)}\left( \lambda \right)=\int_{-\ii\infty}^{\ii\infty}\frac{ds}{2\pi\ii}\,\frac{\Gamma\left(-s \right)}{\Gamma\left(s+n+1\right)}\,\mathcal{J}_k\left(s+\frac{n+p-3}{2}  \right)\left(\frac{\sqrt{\lambda}}{2\pi}\right)^{2s+n}\,.
\label{Ifinal}
\end{align}
In the strong-coupling regime, \(\lambda\to\infty\), the integral receives contributions from the poles on the negative real axis. We therefore close the contour counterclockwise and evaluate the integral by summing the corresponding residues. This reduces the problem to determining the residues of the functions \(\mathcal{J}_k(s)\). From \eqref{J0} we immediately have
\begin{align}
\text{Res}\, \mathcal{J}_0(s)\bigg|_{s=-m} =  (-1)^{m}\frac{(3-2m)\zeta(2m-2)}{(2\pi)^{2m-2}}(1-2^{2m-2})  \equiv f(m)\,.
\end{align}
For generic $k$, from \eqref{recurrence1}-\eqref{recurrence2}, we can derive a recurrence relation for the residues of \(\mathcal{J}_k(s)\) at \(s=-m\), with $m\geq 2$, namely
\begin{align}
&\operatorname{Res}\mathcal{J}_1(s)\big|_{s=-m}+\pi^2\,\operatorname{Res}\mathcal{J}_1(s)\big|_{s=-m-1}
= f(m+1) \,,
\label{recurrenceresidues1} \\
& \operatorname{Res}\mathcal{J}_2(s)\big|_{s=-m}+\pi^2\,\operatorname{Res}\mathcal{J}_2(s)\big|_{s=-m-1}
= \operatorname{Res}\mathcal{J}_1(s)\big|_{s=-m-1}\,, \label{recurrenceresidues2} 
\end{align}
and recursively also for $k\geq 3$. Starting from the initial condition
\begin{align}
\operatorname{Res}\mathcal{J}_1(s)\big|_{s=-1}=0 \,,
\end{align}
 and using \eqref{recurrenceresidues1}-\eqref{recurrenceresidues2}, one can determine the residues at all negative integers. In particular, we get 
\begin{align}
& \text{Res}\, \mathcal{J}_1(s)\bigg|_{s=-m} = \sum_{k=1}^{m-1}(-1)^{m-k-1} \frac{f(k)}{\pi^{2(m-k)}}= \frac{(-1)^m}{\pi^{2m}} \sum_{k=1}^{m-1} (1-2k)\zeta(2k)(2^{-2k}-1) \label{resJ1} \,, \\
& \text{Res}\, \mathcal{J}_2(s)\bigg|_{s=-m} = \sum_{k=1}^{m-1}(-1)^{m-k-1} \frac{(m-k)\,f(k)}{\pi^{2(m-k+1)}}= \frac{(-1)^m}{\pi^{2m+2}} \sum_{k=1}^{m-1} (m-k)(1-2k)\zeta(2k)(2^{-2k}-1) \label{resJ2} \,.
\end{align}
Closed-form expressions can also be obtained for $k\geq 3$, but they are not needed for our purposes. Summing over these residues, we can straightforwardly obtain the large-$\lambda$ behavior of $\mathcal{T}_{k,n}^{(p)}$ up to exponentially suppressed terms. One finds
\begin{align}
&  \mathcal{T}_{0,n}^{(p)}(\lambda) \underset{\lambda \to \infty}{\sim}  \sum_{m=2}^\infty \frac{\Gamma\left( \frac{n+p-3}{2}+m \right)}{\Gamma\left( \frac{n-p+5}{2}-m \right)} \frac{(-1)^m}{\pi^{2m-2}} (3-2m)\zeta(2m-2)(2^{-2m+2}-1)\left( \frac{2\pi}{\sqrt{\lambda}} \right)^{2m+p-3} \label{largeJ0}, \\[0.5em]
&  \mathcal{T}_{1,n}^{(p)}(\lambda) \underset{\lambda \to \infty}{\sim}  \sum_{m=2}^\infty \frac{\Gamma\left( \frac{n+p-3}{2}+m \right)}{\Gamma\left( \frac{n-p+5}{2}-m \right)} \frac{(-1)^m}{\pi^{2m}} \sum_{k=1}^{m-1} (1-2k)\zeta(2k)(2^{-2k}-1)\left( \frac{2\pi}{\sqrt{\lambda}} \right)^{2m+p-3} \label{largeJ1} \,, \\[0.5em]
&  \mathcal{T}_{2,n}^{(p)}(\lambda) \underset{\lambda \to \infty}{\sim}  \sum_{m=2}^\infty \frac{\Gamma\left( \frac{n+p-3}{2}+m \right)}{\Gamma\left( \frac{n-p+5}{2}-m \right)} \frac{(-1)^m}{\pi^{2m+2}} \sum_{k=1}^{m-1} (m-k)(1-2k)\zeta(2k)(2^{-2k}-1) \left( \frac{2\pi}{\sqrt{\lambda}} \right)^{2m+p-3}\label{largeJ2} .
\end{align}
It is worth emphasizing that when $n-p$ is odd, the asymptotic expansions \eqref{largeJ0}-\eqref{largeJ2} truncate after finitely many terms, and may even vanish identically. This is precisely what happens for the cases relevant to our analysis, for which $|n-p|=1$. More explicitly:
\begin{align}
& \mathcal{T}_{0,1}^{(0)}(\lambda) \underset{\lambda \to \infty}{\sim}\frac{\pi}{4\sqrt{\lambda}}\,,\quad \mathcal{T}_{1,1}^{(0)}(\lambda) \underset{\lambda \to \infty}{\sim}\frac{1}{4\pi\sqrt{\lambda}}\,,\quad \mathcal{T}_{1,2}^{(1)}(\lambda) \underset{\lambda \to \infty}{\sim} \frac{1}{2\lambda}\,, \quad \mathcal{T}_{1,3}^{(2)}(\lambda) \underset{\lambda \to \infty}{\sim} \frac{2\pi}{\lambda^{3/2}}\,, \\[0.5em]
& \mathcal{T}_{1,0}^{(1)}(\lambda)\,, \mathcal{T}_{1,1}^{(2)}(\lambda)\,, \mathcal{T}_{2,1}^{(2)}(\lambda)\,, \mathcal{T}_{2,2}^{(3)}(\lambda) \underset{\lambda \to \infty}{\sim} 0 \,.
\label{strongT}
\end{align}
These results are in agreement with those found in \eqref{1t}-\eqref{8t} using a different method.

The fact that the strong-coupling expansions in inverse powers of the ’t Hooft coupling truncate, or even vanish, is fully consistent with the non-trivial dependence of $\mathcal{T}_{k,n}^{(p)}(\lambda)$ on the ’t Hooft coupling. Indeed, in addition to the asymptotic series \eqref{largeJ0}-\eqref{largeJ2}, one must incorporate non-perturbative contributions that are exponentially suppressed in the large-$\lambda$ regime. Such terms cannot be captured by the present method. For this reason, the approach followed in the main text is more powerful: there, the integrals of Bessel functions of the first kind are rewritten as combinations of modified Bessel functions of the second kind, whose asymptotic expansion makes the exponentially suppressed contributions manifest.

\printbibliography

\end{document}